\documentclass[a4paper,11pt]{article}
\pdfoutput=1 % if your are submitting a pdflatex (i.e. if you have
\usepackage{jheppub}
\usepackage{epsfig}
\usepackage{amsmath}
\usepackage{graphicx}
\usepackage{appendix}
\usepackage{titlesec}
\usepackage{subfigure}
\usepackage{subcaption}
\def\A0#1{\Pi_{\rm #1}(0)}
\def\AP0#1{\Pi'_{\rm #1}(0)}
\def\be{\begin{equation}}
\def\ee{\end{equation}}
\def\bea{\begin{array}}
\def\eea{\end{array}}
\def\beqa{\begin{eqnarray}}
\def\eeqa{\end{eqnarray}}
\def\beqas{\begin{eqnarray*}}
\def\eeqas{\end{eqnarray*}}

\def\bp{\begin{picture}}
\def\ep{\end{picture}}
\def\bc{\begin{center}}
\def\ec{\end{center}}
\def\bfig{\begin{figure}}
\def\efig{\end{figure}}

\def\bit{\begin{itemize}}
\def\eit{\end{itemize}}
\def\f{\frac}

\def\[{\left[}
\def\]{\right]}
\def\({\left(}
\def\){\right)}

\def\..{\left.}
\def\.{\right.}

\def\ra{\rightarrow}
\def\la{\leftarrow}

\def\tm{\times}

\def\la{\lambda}

\def\ka{\kappa}

\def\ep{\epsilon}

\def\ga{\gamma}

\renewcommand{\thefootnote}{\fnsymbol{footnote}}

\title{Interpretation of 95 GeV Excess within the Georgi-Machacek Model in Light of Positive Definiteness Constraints}
\author[1,2]{Xiaokang Du,}
\author[1,2]{Huiling Liu}
\author[1,2,3]{Qin Chang\footnotemark[1]\footnotetext[1]{Corresponding author.}}

\affiliation[1]{Centre for Theoretical Physics, Henan Normal University, Xinxiang 453007, P. R. China}
\affiliation[2] {Institute of Physics, Henan Academy of Sciences, Zhengzhou 450046, P. R. China}
\affiliation[3] {Center for High Energy Physics, Henan Academy of Sciences, Zhengzhou 455004, P. R. China}

\emailAdd{xkdu@hnas.ac.cn, lhl@hnas.ac.cn, changqin@htu.edu.cn}

\abstract{The recent observation of a di-photon excess around 95 GeV by the CMS and ATLAS Collaborations, along with the $b\bar{b}$ excess reported by the LEP Collaboration in the same mass region, has drawn significant interest in the possibility of new physics beyond the Standard Model (SM). The Georgi-Machacek (GM) model, which extends the Higgs sector of the SM by introducing additional triplet scalars while preserving custodial symmetry at tree level, provides a compelling framework to explain both excesses simultaneously via a light custodial singlet Higgs. In this work, we investigate whether the GM model can still accommodate these excesses when taking into account newly proposed vacuum stability constraints, particularly the positive definiteness conditions. 
Our numerical analysis not only confirms the existence of a viable parameter space capable of explaining the 95 GeV excesses, but also demonstrates that, compared to traditional tree-level constraints at the electroweak scale, the positive definiteness conditions further expand the allowed parameter space, thereby enhancing the viability of the GM model. Furthermore, we emphasize that future collider experiments will play a crucial role in testing this interpretation by refining Higgs coupling measurements and searching for additional Higgs bosons.}

\begin{document}
\maketitle
\newpage
\section{Introduction}
\label{sec-1}

\renewcommand{\thefootnote}{\textsuperscript{\arabic{footnote}}}

The discovery of the 125 GeV Higgs boson~\cite{ATLAS:higgs,CMS:higgs}, the last elementary particle predicted by the Standard Model (SM), represents a significant milestone in particle physics. Moreover, subsequent electroweak (EW) precision measurements align quite well with the predictions of the SM. However, the remaining theoretical and experimental uncertainties mean that the specific process of electroweak symmetry breaking (EWSB) remains an open question \cite{uncertainties of Higgs sector}. Thus, the search for additional scalar fields beyond the SM is a crucial task at the Large Hadron Collider (LHC).

Recently, both the CMS and ATLAS collaborations reported an intriguing di-photon excess around 95 GeV. CMS observed a di-photon excess of 95.4 GeV with 2.9$\sigma$ local significance based on the full Run-2 data \cite{CMS:95,CMS:95-2}. The corresponding signal strength is $\mu_{\gamma\gamma}^{CMS} = {0.33}_{-0.12}^{+0.19}$. 
Coincidentally, ATLAS also presented a di-photon excess with 1.7$\sigma$ \cite{ATLAS:95} local significance at the same mass region with the signal strength $\mu_{\gamma\gamma}^{ATLAS}={0.18}_{-0.1}^{+0.1}$. When these results are
combined, the local significance rises to 3.1$\sigma$, with a combined signal strength of \cite{combine:95}
\begin{align}
\mu_{\gamma\gamma}^{ATLAS+CMS}={0.24}_{-0.08}^{+0.09}.
\end{align}

Interestingly, search data for a low-mass scalar boson at the Large Electron-Positron (LEP) collider may suggest a local excess with a significance of 2.3$\sigma$ in the  $b\bar{b}$ channel at an invariant mass of approximately 98 GeV  \cite{LEP:95}, accompanied by a signal strength of  
\begin{align}
\mu_{b\bar{b}}^{LEP} = 0.117\pm0.057.
\end{align}
Given the limited mass resolution for the di-jets at LEP, it is plausible that this $b\bar{b}$ excess and di-photon excess may indicate the same scalar particle\footnote{Actually, CMS has also reported di-tau excess at 100 GeV (95 GeV) with local significance of 3.1 $\sigma$ (2.6 $\sigma$) \cite{CMS:95tautau}. While ATLAS has not yet reported any information in the channel around 95 GeV, so we will not discuss the di-tau excess here.} as reported by CMS and ATLAS. 

Given such interesting coincidental signals, it is meaningful to explore whether the final result corresponds to a new light scalar particle from the perspective of new physics. Indeed, many papers have explained these signals within different new physics models  \cite{combine:95,95-1,95-2,95-3,95-4,95-5,95-6,95-7,95-8,95-9,95-10,95-11,95-12,95-13,95-14,95-15,95-16,95-17,95-18,95-19,95-20,95-21,95-22,95-23,95-24,95-25,95-26,95-27,95-28,95-29,95-30,95-31,95-32,95-33,95-34,95-35,95-36,95-37,95-38,95-39,95-40,95-41,95-42,95-43,95-44,95-45,95-46,95-47,95-48,95-49,95-50,95-51,95-52,95-53,95-54,95-55,95-56,95-57,95-58,95-59,95-60,95-61,95-62,95-63,95-64,95-GM-1,95-GM-2,95-GM-3}.

Among these models, the GM model can provide a light custodial singlet scalar as a perfect candidate to interpret di-photon and $b\bar{b}$ excess simultaneously \cite{95-GM-1,95-GM-2,95-GM-3}. There could even be a twin-peak resonance around 95 GeV \cite{95-GM-2}, and demonstrates the nature of di-photon, $b\bar{b}$, and di-tau excesses with two scalars in the GM model (a light $CP$-even scalar and a same-mass $CP$-odd scalar).

The GM model ~\cite{95-GM-1,95-GM-2,95-GM-3,GM,GM2,GM3,GM4,GM5,GM6,GM7,GM8,GM9,GM10,GM11,GM12,GM13,GM14,GM15,GM16,GM17, GM18,GM19,GM20,GM21,GM22,GM23,GM24,GM25,GM26,GM27,GM28,GM29,GM30,GM31,GM32,GM33,GM34,GM35,GM36,GM37,GM38,GM39,GM40,GM41,GM63,GM64,GM66,GM68,GM69,GM70,GM44,GM45,GM46,GM47,GM48,GM49,GM50,GM51,GM52,GM53,GM54,GM55,GM56,GM57,GM58,GM59,GM60,GM61,GM62,GM65,GM67,GM-pd} extends the SM by incorporating a complex $SU(2)_L$ triplet scalar $\chi$ with unit hypercharge and a real $SU(2)_L$ triplet scalar $\xi$ with zero hypercharge.
This intricate construction enables the nontrivial preservation of custodial symmetry at tree level after EWSB, dependent on the precise alignment of the vacuum expectation values (VEVs) between these triplet scalars, which ensures that the triplet state expectation value can reach several tens of GeV, significantly higher than in new physics models that extend the SM with a triplet. In addition to the unique additional scalar sector, the GM model predicts interesting phenomenology  \cite{GM24,GM25,GM26,GM27,GM28,GM29,GM30,GM31,GM32,GM33,GM34,GM35,GM36,GM37,GM38,GM39,GM40,GM41,GM63,GM64,GM66}, such as enhanced couplings of the SM-like Higgs boson to W and Z gauge bosons, as well as the existence of nine more physical scalars, including a doubly charged scalar particle. 

In the GM model related studies, theoretical constraints from perturbative unitarity, the EW vacuum stability and conditions to avoid alternative minima vacuum~\cite{GM57,GM58,GM59,GM60,GM61,GM62,GM65,GM67} etc should be satisfied other than those exclusion bounds from new particle searches~\cite{GM56,GM59} by colliders. In our previous work \cite{GM-pd}, we have rediscussed the vacuum stability  bound for the GM model, concentrating specially on the constraints bounds from non-existence of a deeper vacuum at large field values. The numerical results indicate that quite a lot parameter space satisfied theoretical and experimental constraints have been excluded by the new vacuum stability bounds. It is interesting to explore whether there is still parameter space within the GM model that can accommodate the 95 GeV excess, considering all the theoretical and experimental constraints.

This work is organized as follows. In Sec~\ref{sec-2}, we briefly review the GM model. A general discussion about di-photon and $b\bar{b}$ excess signals at mass of 95 GeV is given in Sec~\ref{sec-3}. In Sec~\ref{sec-4}, we listed the theoretical and experimental constraints in realistic research, and reviewed the bounds of positive definiteness  proposed in our previous work. The numerical results are shown in Sec~\ref{sec-5}. In Sec~\ref{sec-6} we present our conclusion. 

\section{\label{sec-2} A brief review of the Georgi-Machacek model}
%\subsection{\label{sec-21} review of the Georgi-Machacek model}

The GM model's Higgs sector encompasses the well-known SM-like $SU(2)_L$ complex doublet Higgs field, labeled as $({\phi}^+,~{\phi}^0)$, which carries a hypercharge of $1/2$. Furthermore, it incorporates two additional $SU(2)_L$ triplet Higgs fields: one complex triplet, identified as $({\chi}^{++},~{\chi}^{+},~{\chi}^{0})$) with a unit hypercharge, and a real triplet possessing a zero hypercharge, denoted as $({\xi}^{+},~{\xi}^{0},~-{\xi}^{+*})$. This extension introduces novel features and complexities in the Higgs sector compared to the Standard Model. The general potential\footnote{The last two terms in the scalar potential can be absent if introducing a $Z_2$ symmetry.} of GM model being invariant under the global $SU(2)_L \times SU(2)_R$ symmetry and SM gauge symmetry can be given as
\begin{align}
V(\Phi,\Delta) = & \frac12 m_{\Phi}^2 {\rm tr} \left[ \Phi^\dagger \Phi \right] + \frac12 m_{\Delta}^2 {\rm tr} \left[ \Delta^\dagger \Delta \right] + \lambda_1 \left( {\rm tr} \left[ \Phi^\dagger \Phi \right] \right)^2 + \lambda_2 \left( {\rm tr} \left[ \Delta^\dagger \Delta \right] \right)^2 \notag \\
& + \lambda_3 {\rm tr} \left[ \left( \Delta^\dagger \Delta \right)^2 \right] + \lambda_4 {\rm tr} \left[ \Phi^\dagger \Phi \right] {\rm tr} \left[ \Delta^\dagger \Delta \right] + \lambda_5 {\rm tr} \left[ \Phi^\dagger \frac{\sigma^a}{2} \Phi \frac{\sigma^b}{2} \right] {\rm tr} \left[ \Delta^\dagger T^a \Delta T^b \right] \notag \\
& + \mu_1 {\rm tr} \left[ \Phi^\dagger \frac{\sigma^a}{2} \Phi \frac{\sigma^b}{2} \right] (P^\dagger \Delta P)_{ab} + \mu_2 {\rm tr} \left[ \Delta^\dagger T^a \Delta T^b \right] (P^\dagger \Delta P)_{ab},
\label{eq:GMpot}
\end{align}
where the $\Phi$ and $\Delta$ are the matrix form of doublet and triplets to tell whether the custodial symmetry is preserved or not more easily. And the specific representation can be shown as
\begin{align}
\Phi=\left(
\begin{array}{cc}
\phi^{0*} & \phi^+ \\
\phi^- & \phi^0
\end{array}\right),\quad
\Delta=\left(
\begin{array}{ccc}
\chi^{0*} & \xi^+ & \chi^{++} \\
\chi^- & \xi^0 & \chi^{+} \\
\chi^{--} & \xi^- & \chi^{0}
\end{array}\right)~.
\label{eq:Higgs_matrices}
\end{align}
The $\sigma^i~\left(i=1,2,3\right)$  (the Pauli matrices) in the potential are the generators of the $\left(\textbf{2},\bold{\bar{2}}\right)$ representation of $SU(2)$ group, and the $T^i~\left(i=1,2,3\right)$ are the generators for the $\left(\textbf{3},\bold{\bar{3}}\right)$ representation 
\begin{align}
T^1 = & \frac{1}{\sqrt{2}} \left( \begin{array}{ccc} 0 & 1 & 0 \\ 1 & 0 & 1 \\ 0 & 1 & 0 \end{array} \right),~T^2=\frac{1}{\sqrt{2}} \left( \begin{array}{ccc} 0 & -i & 0 \\ i & 0 & -i \\ 0 & i & 0 \end{array} \right),~T^3 = \left( \begin{array}{ccc} 1 & 0 & 0 \\ 0 & 0 & 0 \\ 0 & 0 & -1 \end{array} \right).
\end{align}
Besides, the $P$ is a matrix to rotate the triplet into its Cartesian basis
\begin{align}
P = & \frac{1}{\sqrt{2}} \left( \begin{array}{ccc} -1 & i & 0 \\ 0 & 0 & \sqrt{2} \\ 1 & i & 0 \end{array} \right).
\end{align}

The neutral real scalar field can be parameterized as ${\xi}^0  \ra v_{\xi}+h_{\xi}$, and the neutral complex scalar fields are decomposed into real and imaginary parts as
\beqa
{\phi}^0  \ra \left(v_{\phi}+h_{\phi}+i a_{\phi}\right)/\sqrt{2},~~~{\chi}^0  \ra v_{\chi}+\left(h_{\chi}+i a_{\chi}\right)/\sqrt{2}.
\eeqa 
The parameters $v_{\xi}$, $v_{\phi}$ and $v_{\chi}$ are VEVs of corresponding neutral scalar fields that trigger the breaking of $SU(2)_L \times SU(2)_R$ symmetry. Especially, the custodial $SU(2)_V$ symmetry would be preserved at tree level if we choose alignment VEVs for triplets $v_{\chi}=v_{\xi}=v_{\Delta}$. In other words, VEVs of scalar fields take the identity form
\beqa
\langle\Phi\rangle = \frac{1}{\sqrt{2}}\left( \begin{array}{cc} v_{\phi} & 0\\ 0 & v_\phi \end{array} \right),~\langle\Delta\rangle=\left( \begin{array}{ccc} v_\Delta & 0 & 0 \\ 0 & v_\Delta & 0 \\ 0 & 0 & v_\Delta \end{array} \right).
\eeqa 
 Henceforth the EWSB condition becomes
 \beqa
v^2=v_\phi^2+4 v_\chi^2+4 v_\xi^2=v_\phi^2+8 v_\Delta^2=\f{1}{\sqrt{2}G_F}\approx \(246 {\rm GeV}\)^2~.
\eeqa
The tadpole equations can be obtained by minimizing scalar potential with VEVs 
\begin{gather}
\frac{\partial V}{\partial v_\phi}=0 \ra m_\phi^2=-4\lambda_1 v_\phi^2-6\lambda_4 v_\Delta^2 - 3\lambda_5 v_\Delta^2 - \frac{3}{2}\mu_1 v_\Delta^2, \notag \\
\frac{\partial V}{\partial v_\Delta}=0 \ra m_\Delta^2=-12\lambda_2 v_\Delta^2-4\lambda_3 v_\Delta^2 - 2\lambda_4 v_\phi^2 - \lambda_5 v_\phi^2 - \frac{\mu_1 v_\phi^2}{4 v_\Delta}\mu_1 - 6 \mu_2 v_\Delta~.
\end{gather}

For $\textbf{2}\otimes\bold{2}=\textbf{1}\oplus\textbf{3}$ and $\textbf{3}\otimes\bold{3}=\textbf{1}\oplus\textbf{3}\oplus\textbf{5}$, after $SU(2)_L \times SU(2)_R$ symmetry breaking down to the custodial $SU(2)_V$ symmetry, we can have 2 $CP$-even singlets ($H_1,~H_1^\prime$), 2 $CP$-odd triplets ($G_3,H_3$) and 1 $CP$-even quintuplet ($H_5$) by their transformation properties.
$G_3$ is the so-called Goldstone boson, absorbed by W and Z bosons after EWSB
\beqa
G_3^\pm=\cos{\theta}\phi^\pm+\sin{\theta}\frac{\chi^\pm+\xi^\pm}{\sqrt{2}},~G_3^0=\cos{\theta}\phi^{0,i}+\sin{\theta}\chi^{0,i}~,
\eeqa
where ``i" denotes the imaginary part of the complex scalar fields. The mixing parameter $\theta$ can be obtained through  the relation $\tan{\theta}\equiv2\sqrt{2}v_\Delta/v_\phi$, allowing for a quantitative assessment of the respective contributions made by the doublet and triplets to EWSB.

The presence of the tree-level custodial symmetry permits the VEVs $v_\Delta$ of the triplets to attain values as substantial as 60 GeV, when solely considering constraints imposed by perturbativity, vacuum stability, and indirect limits derived from oblique parameters and Z-pole observables, among others. However, when incorporating collider constraints on the scalar states as reported in Ref. \cite{GM32}, the upper limit on $v_\Delta$ undergoes a reduction to 40 GeV, necessitating that the mixing angle $\theta$ associated with the triplet contribution be confined to a value less than 0.45.

The physical triplet is
\beqa
H_3^\pm=\cos{\theta}\frac{\chi^\pm+\xi^\pm}{\sqrt{2}}-\sin{\theta}\phi^\pm,~H_3^0=\cos{\theta}\chi^{0,i}-\sin{\theta}\phi^{0,i}~.
\eeqa
The quintuplet is organized as 
\beqa
H_5^{\pm\pm}=\chi^{\pm\pm},~ H_5^{\pm}=\frac{\chi^\pm-\xi^\pm}{\sqrt{2}},~H_5^0=\sqrt{\frac{2}{3}}\xi^0-\sqrt{\frac{1}{3}}\chi^{0,r}~,
\eeqa
where the ``r" means the real part of complex scalar field.
The degenerate masses at tree level of triplet and quintuplet can be given as
\beqa
M_3^2=\frac{m_\Phi}{4 v_\Delta}v^2+\frac{\lambda_5}{2}v^2,~M_5^2=\frac{m_\Phi}{4 v_\Delta}v_\phi^2+12 M_\Delta v_\Delta+\frac{3}{2}\lambda_5 v_\phi^2+8\lambda_3 v_\Delta^2,
\eeqa

The two singlets can be given by
\beqa
H_1=\phi^{0,r},~H_1^\prime=\sqrt{\frac{2}{3}}\chi^{0,r}+\sqrt{\frac{1}{3}}\xi^{0},
\eeqa
while this two singlets at last will mix into two physical singlet states
\beqa
\left(h, ~H \right) \left( \begin{array}{cc} m_h^2 & 0\\ 0 & m_H^2 \end{array} \right) \left( \begin{array}{c} h\\ H \end{array} \right) = \left(H_1,~H_1^\prime \right) \left( \begin{array}{cc} \mathcal{M}_{11}^2 & \mathcal{M}_{12}^2\\ \mathcal{M}_{21}^2 & \mathcal{M}_{22}^2 \end{array} \right) \left( \begin{array}{c} H_1\\ H_1^\prime \end{array} \right).
\eeqa
The elements of the mass matrix can be shown as
\begin{gather}
\mathcal{M}_{11}^2=8\lambda_1 v_\phi^2, \notag \\  
\mathcal{M}_{12}^2=\mathcal{M}_{21}^2=\frac{\sqrt{3}}{2} v_\phi \left(4\left(2\lambda_2+\lambda_5\right)v_\Delta+\mu_1\right), \notag \\ 
\mathcal{M}_{22}^2=8\left(\lambda_3 +4\lambda_4\right)v_\Delta^2+6\mu_2 v_\Delta - \frac{\mu_1 v_\phi^2}{4 v_\Delta}.
\label{eq:singlets mass matrix}
\end{gather}

The physical singlet masses can be given by calculating the eigenvalues of this mass matrix
\beqa
m_{h,H}^2=\frac{\mathcal{M}_{11}^2+\mathcal{M}_{22}^2\mp\sqrt{{\left(\mathcal{M}_{11}^2-\mathcal{M}_{22}^2\right)}^2+4{\left(\mathcal{M}_{12}^2\right)^2}}}{2}.
\label{singlets-mass}
\eeqa
The lighter singlet scalar would be a good candidate to accommodate the 95 GeV excess if the heavier one is SM-like Higgs with a mass of 125 GeV.

%%%%%%%%%%%%
\section{\label{sec-3} 95 GeV excess at the GM Model}

\subsection{\label{sec-31} Brief review of the 95 GeV excess}

The CMS and ATLAS collaborations have thoroughly investigated low mass scalar resonance in the di-photon final state which has played a key role in the discovery of the SM-like Higgs boson. Recently, CMS reported a resonance at 95.4 GeV with a local significance of 2.9 $\sigma$ based on  the Run-2 full dataset collected at 13 TeV (an integrated luminosity of $132.2~{fb}^{-1}$)~\cite{CMS:95}, the signal strength being
\begin{align}
\mu_{\gamma\gamma}^{CMS}\equiv\frac{\sigma^{exp}(gg\rightarrow\phi\rightarrow\gamma\gamma)}{\sigma^{SM}(gg\rightarrow h\rightarrow\gamma\gamma)} = {0.33}_{-0.12}^{+0.19}.
\end{align}
The $\sigma^{exp}$ denotes cross section obtained by experiment at the di-photon resonance mass, and $\sigma^{SM}$ denotes the cross section for Higgs boson of the SM at 95 GeV.
Thanks to the new analysis strategy, the signal strength of this new result showed a significant reduction compared to their previous results with $\mu_{\gamma\gamma, 95 GeV}^{CMS}=0.6\pm0.2$ based on the dataset from Run 1 at 8 TeV and the part Run 2 data at 13 TeV (an integrated luminosity of $19.7~{fb}^{-1}$ and $35.9~{fb}^{-1}$)~\cite{CMS:95-2}.

%%%%%%%%%%%%%%%%%%%%%%%%%%%%%%%%%%%%%
In the same mass region, ATLAS collaboration detected a di-photon resonance signal with a local significance of 1.7 $\sigma$ based on the full dataset from Run 2 at 13 TeV~\cite{ATLAS:95}, and the slightly smaller signal strength is given by ATLAS
\begin{align}
\mu_{\gamma\gamma}^{ATLAS}\equiv\frac{\sigma^{exp}(gg\rightarrow\phi\rightarrow\gamma\gamma)}{\sigma^{SM}(gg\rightarrow h\rightarrow\gamma\gamma)} = {0.18}_{-0.1}^{+0.1}.
\end{align}
By combining their results, the local significance can reach 3.1 $\sigma$~\cite{combine:95}, and the signal strength is adjusted to
\begin{align}
\mu_{\gamma\gamma}^{ATLAS+CMS}\equiv\frac{\sigma^{exp}(gg\rightarrow\phi\rightarrow\gamma\gamma)}{\sigma^{SM}(gg\rightarrow h\rightarrow\gamma\gamma)} = {0.24}_{-0.08}^{+0.09}.
\end{align}

Coincidentally, the results of LEP collaboration showed a mild excess with local significance of 2.3$\sigma$ in the $b\bar{b}$ channel around 98 GeV~\cite{LEP:95}, and signal strength being
\begin{align}
\mu_{b\bar{b}}^{LEP} = 0.117\pm0.057.
\end{align}
In view of the limitation of di-jet mass resolution in the LEP, the hadronic $b\bar{b}$ excess may suggest the possibility of a scalar particle at 95 GeV, which is in line with the observed di-photon excess reported by CMS and ATLAS.

%Besides, CMS has identified an excess in the di-tau channel that is consistent with a mass of 95 GeV. The peak of this excess was observed at a mass of 100 GeV, with a local significance of 3.1σ. At 95 GeV, the local significance was found to be 2.6σ. This anomaly could be indicative of a scalar resonance, with a signal strength estimated at µ(Φ95)ττ = 1.23+0.61 −0.49. It should be pointed out that the mass resolution in the di-tau channel is much broader than that of the diphoton channel, which allows for the possibility that a scalar resonance at 95 GeV could be the source of both the diphoton and ditau excesses, despite the fact that the ditau excess is most pronounced at a mass of 100 GeV. While this 

%Whereas, ATLAS has not reported any analogous searches for scalar resonances decaying into a pair of tau leptons in the mass region around 95 GeV.

\subsection{\label{sec-32} The 95 GeV excess at the GM model}
%%%%%%%%%%%%%%%%%%%%%%%%%%%%%%%%%%%%%%%%
The simultaneous detection of a di-photon excess at ATLAS and CMS is quite intriguing, particularly as it both occurs in the vicinity of 95 GeV. The coincidental observation of a $b \bar{b}$ excess in the same mass window, as noted in LEP experiments, adds an additional layer of fascination. Consequently, it becomes crucial to investigate these excesses using new physics models.
Assuming the existence of a new scalar particle, the signal strength can be redefined as 
\begin{align}
\mu(\Phi_{95\mathrm{GeV}})\equiv\frac{\sigma({\Phi_{95\mathrm{GeV}}})\times BR(\Phi_{95\mathrm{GeV}}\rightarrow \gamma\gamma/b\bar{b})}{\sigma_{\mathrm{SM}}({\Phi_{95\mathrm{GeV}}})\times BR_{\mathrm{SM}}(\Phi_{95\mathrm{GeV}}\rightarrow \gamma\gamma/b\bar{b})}.
\end{align}

The GM model predicts two light custodial singlet scalar fields. The masses of them can be accurately estimated by the equation (\ref{singlets-mass}) at the lowest order. The lighter field, referred to as ``$H$", is obviously a natural candidate to explain the observed signals if its mass is approximately 95 GeV, while the heavier field ``$h$" should correspond to the SM-like Higgs boson with a mass around 125 GeV. In the narrow width approximation, the di-photon and $b\bar{b}$ excess signal strengths can be given
\begin{gather}
\mu_{\gamma\gamma}(H)\equiv\frac{\sigma_{\mathrm{GM}}(gg\rightarrow H)\times BR(H\rightarrow \gamma\gamma)}{\sigma_{\mathrm{SM}}(gg\rightarrow h_{\mathrm{SM}})\times BR(h_{\mathrm{SM}}\rightarrow \gamma\gamma)}=\kappa^2_{Hff}\times\frac{BR(H\rightarrow \gamma\gamma)}{BR(h_{\mathrm{SM}}\rightarrow \gamma\gamma)}, \notag \\
\mu_{b\bar{b}}(H)\equiv\frac{\sigma_{\mathrm{GM}}(e^+ e^- \rightarrow ZH)\times BR(H\rightarrow b\bar{b})}{\sigma_{\mathrm{SM}}(e^+ e^- \rightarrow Zh_{\mathrm{SM}})\times BR(h_{\mathrm{SM}}\rightarrow b\bar{b})}=\kappa^2_{HVV}\times\frac{BR(H\rightarrow b\bar{b})}{BR(h_{\mathrm{SM}}\rightarrow b\bar{b})},
\end{gather}
where the $\kappa_{i}$ parameters are the effective Higgs coupling, defined as the square of the ratio of the couplings of the Higgs bosons to fermions (for $i=Hff$) or vector bosons (for $i=HVV$) in new physics models, such as the GM model, relative to their values in the SM. For the lighter scalar $H$, the $\kappa_{i}$ parameters can be denoted as
\begin{gather}
\kappa_{Hff}=g_{H_{\mathrm{{GM}}ff}}/g_{h_{\mathrm{SM}}ff},~~~\kappa_{HVV}=g_{H_{\mathrm{GM}}VV}/g_{h_{\mathrm{SM}}VV}.
\label{kappa}
\end{gather}

In the investigation of resonances around 95 GeV, the choice for $ff$ should be $t\bar{t}$, and for $VV$, it should be $ZZ$ (these choices are due to the requirements of the corresponding resonance production modes at both LHC and LEP). These selections correspond to the production modes that are relevant to both LEP and LHC experiments. For the GM model, the effective Higgs couplings to $t\bar{t}$ and $ZZ$ in the lowest order can be shown as
\begin{gather}
\kappa_{Ht\bar{t}}=\frac{\sin{\alpha}}{\cos{\theta}},~~~\kappa_{ht\bar{t}}=\frac{\cos{\alpha}}{\cos{\theta}}, \notag \\
\kappa_{HZZ}=\cos{\theta}\sin{\alpha} + \sqrt{\frac{8}{3}}\sin{\theta}\cos{\alpha},~~~\kappa_{hZZ}=\cos{\theta}\cos{\alpha} - \sqrt{\frac{8}{3}}\sin{\theta}\sin{\alpha},
\end{gather}
where $\alpha$ is the mixing angle of the two singlets $H_1$ and $H_1^\prime$, denoted as
\begin{align}
\tan{2\alpha}=\frac{2 M_{12}^2}{M_{22}^2-M_{11}^2}.
\label{eq:singlet mixing angle}
\end{align}
It is easy to see that the effective Higgs coupling of the SM-like Higgs $\kappa_{ht\bar{t}/ZZ}$ will approach 1 if the values of $\alpha$ and $v_\Delta$ are very small, as anticipated. Sufficient mixing between the singlets is crucial for addressing the $b \bar{b}$ excess. Additionally, charged Higgs bosons ($H_3^{\pm}$ and $H_5^{\pm}$) and the doubly-charged Higgs boson ($H_5^{\pm\pm}$) play a significant role in the loop-induced di-photon decay model, contributing positively to the interpretation of the di-photon excess signal, beyond the contributions from the W boson and top quark. Both the effective Higgs coupling and the branching ratios at the loop level can be calculated using spectrum generator package.

%%%%%%%%%%%%%%%%%%%%%%%%%%%%%%%%%%%%%%%%%%%%%%%
\section{\label{sec-4} Constraints on the GM Model}
\subsection{\label{sec-41} The Usual Constraints}
In practical research, the input parameters for a realistic GM model must satisfy not only the complex constraints imposed by current experimental data but also adhere to fundamental theoretical requirements. Here are the usual constraints that have to be met. 
\begin{itemize}
    \item 
 \textbf{The perturbative unitarity bounds: }
 The quartic coupling parameters within the scalar potential must conform to the perturbative unitarity bounds derived from the S-wave amplitudes for elastic scattering of two scalar bosons, as detailed in references~\cite{GM58,GM65,Aoki:2007ah}. These bounds are specified by the following
 \begin{eqnarray}
	\sqrt{(6 \lambda_1 -7 \lambda_3 -11 \lambda_4)^2 +36 \lambda_2^2} + \left| 6 \lambda_1 + 7 \lambda_3 +11 \lambda_4 \right| &<& 4 \pi, \notag \\
	\sqrt{(2 \lambda_1 + \lambda_3 -2 \lambda_4)^2 + \lambda_5^2} + \left| 2 \lambda_1 - \lambda_3 +2 \lambda_4 \right| &<& 4 \pi, \notag \\
	\left| 2 \lambda_3 + \lambda_4 \right| &<& \pi, \notag \\
	\left| \lambda_2 - \lambda_5 \right| &<& 2 \pi.
	\label{eqn:pertUnitarity}
 \end{eqnarray}

    \item 
 \textbf{The vacuum stability constraints: } 
 Current vacuum stability constraints are simply based on the requirement that the tree-level scalar potential must be bounded from below~\cite{GM58}. The quadratic and cubic terms in the potential~(\ref{eq:GMpot}) can be safely disregarded because the quartic terms will dominate the tree-level scalar potential at large field values. Thus the custodial scalar potential (\ref{eq:GMpot}) can thus be reformulated as
 \begin{equation}
	V^{(4)}(r, \tan \gamma, \zeta , \omega ) = \left[ \lambda_1 + (\lambda_2 - \omega \lambda_5) \tan^2 \gamma 
	+ (\zeta \lambda_3 + \lambda_4 ) \tan^4 \gamma \right] \frac{r^4}{(1+\tan^2 \gamma)^2},
	\label{eq:bfb}
\end{equation}
with the following definitions
\begin{eqnarray}
	r &\equiv&  \sqrt{\text{Tr}(\Phi^\dagger \Phi) + \text{Tr}({\Delta}^\dagger {\Delta})} ,  \notag \\
	r^2 \cos^2 \gamma  &\equiv&  \text{Tr}(\Phi^\dagger \Phi)   ,  \notag \\   
	r^2 \sin^2 \gamma &\equiv&   \text{Tr}({\Delta}^\dagger {\Delta})   ,    \notag \\    
	\zeta &\equiv&   \frac{ \textrm{Tr}({\Delta}^\dagger {\Delta} {\Delta}^\dagger {\Delta})}{[{\rm Tr}({\Delta}^\dagger {\Delta})]^2},  \notag \\
	\omega &\equiv&   \frac{\text{Tr}( \Phi^\dagger \frac{\sigma^a}{2} \Phi \frac{\sigma^b}{2}) \text{Tr}( {\Delta}^\dagger T^a {\Delta} T^b)}{\textrm{Tr}(\Phi^\dagger \Phi)\textrm{Tr}({\Delta}^\dagger {\Delta})}.
	\label{eq:bfbdefs}
\end{eqnarray}
And the ranges of these parameters are ~\cite{GM58}
\begin{equation}
 	r \in [0,\infty)  ,  \quad   
	\gamma \in \left[ 0,\frac{\pi}{2} \right]   ,    \quad  
	\zeta \in \left[ \frac{1}{3},1 \right]  \quad  \textrm{and}  \quad  
	\omega \in \left[ -\frac{1}{4}, \frac{1}{2} \right].
 \end{equation}  

Thus, the requirement of the bounded-from-below conditions of the tree-level scalar potential in this form establishes the necessary and sufficient conditions that the quartic couplings have to obey
\begin{eqnarray}
	\lambda_4 &>& \left\{ \begin{array}{l l}
		- \frac{1}{3} \lambda_3 & {\rm for} \ \lambda_3 \geq 0, \\
		- \lambda_3 & {\rm for} \ \lambda_3 < 0, \end{array} \right. \notag \\
	\lambda_2 &>& \left\{ \begin{array}{l l}
		\frac{1}{2} \lambda_5 - 2 \sqrt{\lambda_1 \left( \frac{1}{3} \lambda_3 + \lambda_4 \right)} &
			{\rm for} \ \lambda_5 \geq 0 \ {\rm and} \ \lambda_3 \geq 0, \\
		\omega_+(\zeta) \lambda_5 - 2 \sqrt{\lambda_1 ( \zeta \lambda_3 + \lambda_4)} &
			{\rm for} \ \lambda_5 \geq 0 \ {\rm and} \ \lambda_3 < 0, \\
		\omega_-(\zeta) \lambda_5 - 2 \sqrt{\lambda_1 (\zeta \lambda_3 + \lambda_4)} &
			{\rm for} \ \lambda_5 < 0, 
			\end{array} \right.  \notag \\
        \lambda_1 &>& 0,
	\label{eq:bfbcond2}
\end{eqnarray}
where the value of $\omega_{\pm}(\zeta)$ for any $\zeta$ can be calculated with  
\begin{equation}
\omega_{\pm}(\zeta) = \frac{1}{6}(1 - \sqrt{\frac{3}{2}\left(\zeta - \frac{1}{3}\right)}) \pm \frac{\sqrt{2}}{3} \left[ (1 - \sqrt{\frac{3}{2}\left(\zeta - \frac{1}{3}\right)}) \left(\frac{1}{2} + \sqrt{\frac{3}{2}\left(\zeta - \frac{1}{3}\right)}\right)\right]^{1/2}.
\end{equation}

    \item 
 \textbf{Relevant constraints about Higgs: } 
 The mass of the light singlet scalar $H$ is anticipated to be around 95.4 GeV to explain the observed di-photon and $b\bar{b}$ excesses, with a theoretical and experimental uncertainty of $±1$ GeV, setting the mass range from 94.4 GeV to 96.4 GeV. The properties of the heavier singlet scalar $h$, corresponding to the SM-like Higgs boson found at the LHC, should match ATLAS and CMS measurements at the $95\%$ confidence level, requiring p-value greater than 0.05 and $\Delta_{\chi^2}=\chi^2_{total}-\chi^2_{SM}$ smaller than 6.18, as verified by the package of HiggsSignal-2.6.2 \cite{HiggsSignals1,HiggsSignals2,HiggsSignals3,HiggsTools}. Moreover, the signal rates for extra Higgs bosons must be consistent with cross-section limits from collider data (including LHC, LEP, etc), which can be checked by the package of HiggsBounds-5.10 \cite{HiggsTools,HiggsBounds1,HiggsBounds2,HiggsBounds3,HiggsBounds4}.

    \item 
 \textbf{Electroweak precision measurements: } 
 The latest global fit results for the Peskin-oblique parameters, based on the precision electroweak data, can be presented as
 \cite{PDG:2022}
 \beqa
 S = 0.02 \pm 0.10,~T = 0.03\pm 0.12, ~U = 0.01\pm 0.11,
 \eeqa
 and the correspond electroweak $\rho$ parameter should be
 \beqa
 \rho = 1.00038 \pm 0.00020.
 \eeqa
 This so tiny uncertainty is an interesting sign to the GM model as candidate of new physics for its preservation of the custodial symmetry at tree level.
 
 As noted in eq(\ref{kappa}), the couplings of SM-like Higgs $h$ to the SM fermions and weak gauge bosons $V = W, Z$ are modified in GM-type models, and the deviations from the SM predictions can be parameterized by the so-called ``effective Higgs coupling"$ \ka_{ff}$ and $\ka_{VV}$.
 The most strict constraints are given by ATLAS and CMS collaborations \cite{ATLAS:kappaWZ,CMS:kappaWZ,PDG:2022}
 \begin{gather}
 \ka_{tt}=0.94\pm 0.11,~\ka_{bb}=0.89\pm 0.11,~\ka_{\tau\tau}=0.93\pm 0.07, \notag \\
 \ka_{\mu\mu}=1.06^{+0.25}_{-0.30},~\ka_{WW}=1.05\pm 0.06,~\ka_{ZZ}=0.99\pm 0.06,~~({\rm ATLAS})~,
 \label{ATLAS:kappa}
 \end{gather}
 and 
 \begin{gather}
 \ka_{tt}=1.01\pm 0.11,~\ka_{bb}=0.99\pm 0.16,~\ka_{\tau\tau}=0.92\pm 0.08, \notag \\
 \ka_{\mu\mu}=1.12\pm 0.21,~\ka_{WW}=1.02\pm 0.08,~\ka_{ZZ}=1.04\pm 0.07,~~({\rm CMS})~.
 \label{CMS:kappa}
 \end{gather}
 Thus, we would like our parameter space to range within
 \begin{gather}
 0.83\leq\ka_{tt}\leq 1.10,~0.78\leq\ka_{bb}\leq1.15,~0.86\leq\ka_{\tau\tau}\leq1.00, \notag \\
 0.76\leq\ka_{\mu\mu}\leq1.33,~0.94\leq\ka_{WW}\leq1.11,~0.93\leq\ka_{ZZ}\leq1.11.
 \label{constraints of kappa}
 \end{gather}
 %It appears that the effective Higgs coupling $\ka_{ff}$ tends to be less than 1, whereas the $\ka_{VV}$ tend to be slightly greater than 1. This characteristic is a notable feature of the GM model and adds another intriguing layer to our understanding of it.
 
 \item
 \textbf{B-physics observables: } 
 The rare decays of B-mesons are highly sensitive probes for new physics because these decay processes are typically very suppressed in the SM. Consequently, the observables from B-meson rare decays often serve as crucial tools for detecting new physics. Here are some common constraints on B-meson rare decays in our realistic research \cite{B-physics}
 \begin{gather}
  1.7\tm 10^{-9} < BR(B^0_s\ra \mu^+ \mu^-) < 4.5\tm 10^{-9}~, \notag \\
  0.85\tm 10^{-4} < BR(B\ra \tau\nu) < 2.89\tm 10^{-4}~, \notag \\
  2.99\tm 10^{-4} < BR(B\ra X_s \gamma) < 3.87\tm 10^{-4}~.
 \end{gather}
\end{itemize}

%%%%%%%%%%%%%%%%%%%%%%%%%%%%%%%%%%%%

\subsection{\label{sec-42} Positive Definiteness Constraints}

The existing constraints on bounded-from-below are derived with the scalar potential at the tree level. It is recognized that quantum corrections have the potential to alter the scalar potential's shape, which might result in the violation of these positive definiteness conditions. This could lead to new deeper vacua at large field values or new vacua that are even not bounded from below, which means that the electroweak vacuum that preserves custodial symmetry might not be the true vacuum similar to the metastable vacuum for a 125 GeV Higgs boson in the SM.

In light of these considerations, we have introduced the so-called ``Positive Definiteness Constraints" within the GM model to ensure the stability of the custodial vacuum \cite{GM-pd}. Here we briefly review the Positive Definiteness Constraints on the GM model.

\begin{itemize}
\item \textbf{Effective potential in the GM model}

For the Higgs sector of the GM model, the full effective potential is constituted by the summation of the classical potential and the quantum loop corrections. The effective scalar potential can be very well approximated by the Renormalization Group improved (RG-improved) tree-level expression in the large field value regions with ($\Delta \gg v_{\Delta},\Phi\gg v_\phi$) after removing the Goldstone combinations of scalars 
\begin{align}
V(\Phi,\Delta) &=\lambda_1(\mu) \left( {\rm tr} \left[ \Phi^\dagger \Phi \right] \right)^2 + \lambda_2(\mu) \left( {\rm tr} \left[ \Delta^\dagger \Delta \right] \right)^2 \notag  + \lambda_3(\mu) {\rm tr} \left[ \left( \Delta^\dagger \Delta \right)^2 \right] \\& + \lambda_4(\mu) {\rm tr} \left[ \Phi^\dagger \Phi \right] {\rm tr} \left[ \Delta^\dagger \Delta \right] + \lambda_5(\mu) {\rm tr} \left[ \Phi^\dagger \frac{\sigma^a}{2} \Phi \frac{\sigma^b}{2} \right] {\rm tr} \left[ \Delta^\dagger T^a \Delta T^b \right] \notag\\
&+ V_{{c-v}}
~,
\label{eq:RGtree}
\end{align}
where the energy scale $\mu$ is at the order of ${\cal O}(h_{\phi;0},h_{\chi;0},h_{\xi;0},\cdots)$. 
$V_{c-v}$ represents the cumulative effect of terms arising from quantum effects that break the  custodial symmetry. Especially, in the direction of  $\Delta$ (or $\Phi$) taking large field values with not very large $\Phi$ (or $\Delta$), it's reasonable to take the form ($\ref{eq:RGtree}$) with the corresponding $\Phi$ (or $\Delta$) terms neglected. 

In the GM model, the quartic coupling coefficients  may change signs at large field values, analogous to what occurs when the quartic coupling $\lambda$ turns negative around ${10}^{11}$ GeV in the SM, potentially leading to the emergence of new, deeper vacua at large field values. These effects may pose a threat to the positive definiteness of the effective scalar potential. Therefore,
it's necessary and sufficient to use the RG-improved tree-level scalar potential to determine whether deeper vacua appear in various regions of large field values.

It is well known that the custodial symmetry  preserved by the GM scalar potential will be violated by quantum corrections, for example, loops involving  $U(1)_Y$ gauge boson. Therefore, it is insufficient to adopt the Lagrangian with custodial symmetry preserving  scalar potential for a proper Renormalization Group Evolution (RGE) description, as new custodial symmetry breaking terms will be generated so that divergent loop corrections related to such terms can not be redefined properly. Therefore, one should begin with the most general $SU(3)_C \times SU(2)_L \times U(1)_Y$ gauge invariant scalar potential~\cite{GM44,GM45}, which includes explicitly all possible custodial symmetry violation terms that can be consistent with the gauge symmetry. 

The custodial symmetry preserved by the GM scalar potential can be broken by quantum corrections, such as those from loops involving the $U(1)_Y$ gauge boson, new terms that break custodial symmetry generated during the process of RGE, can lead to divergent loop corrections. Consequently, it is imperative to start with the most general $SU(3)_C \times SU(2)_L \times U(1)_Y$ gauge invariant scalar potential  that includes all possible terms that violate custodial symmetry while remaining consistent with the gauge symmetry~\cite{GM44,GM45}, which can be written as
\footnote{We are making the assumption that both  $\mu_2$ and $\sigma_4$  are real numbers in this context.}

\begin{align}
V(\phi,\chi,\xi)&=m_\phi^2(\phi^\dagger \phi)+m_\chi^2\text{tr}(\chi^\dagger\chi)+m_\xi^2\text{tr}(\xi^2)\notag\\
&+\mu^\prime_1\phi^\dagger \xi\phi +\mu^\prime_2 [\phi^T(i\tau_2) \chi^\dagger \phi+\text{h.c.}] +\mu^\prime_3\text{tr}(\chi^\dagger \chi \xi) \notag\\
&+\lambda (\phi^\dagger \phi)^2 \notag+\rho_1[\text{tr}(\chi^\dagger\chi)]^2+\rho_2\text{tr}(\chi^\dagger \chi\chi^\dagger \chi)
+\rho_3\text{tr}(\xi^4)\\
&+\rho_4 \text{tr}(\chi^\dagger\chi)\text{tr}(\xi^2)+\rho_5\text{tr}(\chi^\dagger \xi)\text{tr}(\xi \chi)\notag+\sigma_1\text{tr}(\chi^\dagger \chi)\phi^\dagger \phi \notag\\
&+\sigma_2 \phi^\dagger \chi\chi^\dagger \phi
+\sigma_3\text{tr}(\xi^2)\phi^\dagger \phi+\sigma_4 (\phi^\dagger \chi\xi (i\tau_2)\phi^* + \text{h.c.}),
\label{eq:GMpot_full}
\end{align}
where
\begin{align}
\phi = \left(
\begin{array}{c}
\phi^+ \\
\phi^0
\end{array}\right),~
\chi = \left(
\begin{array}{cc}
\frac{\chi^+}{\sqrt{2}} & -\chi^{++}\\
\chi^0 & -\frac{\chi^+}{\sqrt{2}}
\end{array}\right),~
\xi = \left(
\begin{array}{cc}
\frac{\xi^0}{\sqrt{2}} & -\xi^+\\
-\xi^- & -\frac{\xi^0}{\sqrt{2}}
\end{array}\right). \label{par1}
\end{align}

The scalar potential that preserves the custodial symmetry of the original GM model, as shown in equation(\ref{eq:GMpot}), can be derived by identifying specific conditions
\begin{eqnarray}
m_\phi^2&=&2m_\Phi^2,~ m_\chi^2=2m_\Delta^2,~ m_\xi^2= m_\Delta^2,~
\mu^\prime_1=-\frac{\mu_1}{\sqrt{2}},~ \notag \\
\mu^\prime_2&=&-\frac{\mu_1}{2},~
\mu^\prime_3=6\sqrt{2}\mu_2,
\lambda=4\lambda_1,~ 
\rho_1=4\lambda_2+6\lambda_3,~\notag \\ 
\rho_2&=&-4\lambda_3,~ 
\rho_3=2(\lambda_2+\lambda_3),~
\rho_4=4\lambda_2,~
\rho_5=4\lambda_3,~\notag \\
\sigma_1 &=& 4\lambda_4-\lambda_5,~ 
 \sigma_2 =2\lambda_5,~ 
 \sigma_3 =2\lambda_4,~
 \sigma_4 =\sqrt{2}\lambda_5. 
\label{rel} 
\end{eqnarray}
%The tree-level scalar potential (\ref{eq:GMpot}) can preserves the custodial symmetry, t
The deviations in each equation of~(\ref{rel}) during the RGE reflects the generation of custodial symmetry breaking terms in the effective scalar potential from loops involving $U(1)_Y$ gauge interactions. 

A comprehensive analysis of the effects of custodial symmetry breaking in the GM model at higher energies can be found in Ref.~\cite{GM44}, assuming that no custodial symmetry-violating terms are present at the EW scale. The study demonstrates that, within certain parameter regions, the influence of custodial symmetry breaking can be adequately controlled up to the grand unified theory (GUT) scale.
Alternatively, if custodial symmetry is preserved at high-energy scales within a UV completion of the GM model, the measured value of the electroweak $\rho$  parameter, along with the condition of perturbative unitarity, places stringent limits on the scale of this ultraviolet completion across almost the entire parameter space~\cite{GM45}.

Despite their different approaches, the two methods used to explore the custodial symmetry breaking effects induced by RGE between various energy scales yield equivalent consequences in the frame of the GM model.
And one-loop Renormalization Group Equations (RGEs) for scalar quartic couplings $\lambda,~\rho_{1,2,3,4,5},~\sigma_{1,2,3,4}$, the gauge couplings $g_i (i=1,2,3)$, and the Yukawa couplings $Y_{t,b,\tau}$, etc. have been provided in \cite{GM44,GM45}, with two-loop RGEs also available in \cite{GM65}, which can be used to conduct the RGE analysis. 
In practical numerical analyses, the Mathematica package SARAH \cite{SARAH1,SARAH2,SARAH3,SARAH4,SARAH5} not only facilitates the generation of source codes for the spectrum-generator package SPheno \cite{SPheno1,SPheno2} but also can be utilized for RGE numerical studies.

\item \textbf{Positive definiteness of the renormalization group improved tree-level scalar potential}

We utilize the one-loop RG-improved tree-level scalar potential to numerically analyze the vacuum structure and the constraints of positive definiteness in the GM model, which ensures that no deeper vacuum states can emerge across various large field value regions.
To confirm the positive definiteness of the effective scalar potential within the GM model, or to rule out the existence of alternative minima in the large field value regions (where the EWSB and custodial preserving minimum is indeed the true global minimum), it suffices to verify that the positive definite bounds for the scalar potential are maintained. This can be achieved by substituting all corresponding fields with their VEVs and incorporating the relevant minimization conditions.

The invariance of the $SU(2)_L$ gauge symmetry allows us to rotate away three redundant VEVs of the four components of weak isospin components, leaving only $v_\phi$. Furthermore, the invariance of the $SU(2)_V$ custodial symmetry can be used to render three of the VEVs within the triplets redundant, so we can choose to eliminate the VEVs for $\chi^{++}$. Besides, considering any attempt of spontaneous $CP$ breaking would lead to the breaking of the custodial symmetry at EW scale \cite{GM60}, we can assume $CP$ conservation ($v_{\chi;+}^I=v_{\chi;0}^I=0$) in this work. We are left with the following form of the VEVs after denoting $v_{\chi;+}^R\equiv v_{\chi;+}$ and $v_{\chi;0}^R=v_{\chi;0}$ for simplicity
\begin{align}
\langle\phi\rangle = \left(
\begin{array}{c}
0 \\
\f{v_\phi}{\sqrt{2}}
\end{array}\right),~
\langle\chi\rangle = \left(
\begin{array}{cc}
\frac{v_{\chi;+}}{\sqrt{2}} & 0\\
v_{\chi;0} & -\frac{v_{\chi;+}}{\sqrt{2}}
\end{array}\right),~
\langle\xi\rangle = \left(
\begin{array}{cc}
\frac{v_{\xi;0}}{\sqrt{2}} & -v_{\xi;+}\\
-v_{\xi;+} & -\frac{v_{\xi;0}}{\sqrt{2}}
\end{array}\right). \label{par1}
\end{align}

Therefore the RG-improved scalar potential~(\ref{eq:GMpot_full}) at the tree level can be given approximately as
\beqa
V(v_\phi,v_\chi,v_\xi)&=&\f{\la}{4} v_\phi^4+
\rho_1\(v_{\chi;+}^2+v_{\chi;0}^2\)^2+\rho_2\(\f{1}{2}v_{\chi;+}^4+v_{\chi;0}^4+3v_{\chi;+}^2v_{\chi;0}^2\)\notag
\\
&+&2{\rho_3}\(\f{1}{4}v_{\xi;0}^4+v_{\xi;+}^4+v_{\xi;0}^2v_{\xi;+}^2\) +\rho_4\(v_{\chi;+}^2+v_{\chi;0}^2\)\(2v_{\xi;+}^2+v_{\xi;0}^2\)\notag\\
&+&\rho_5\(v^2_{\xi;0}v^2_{\chi;+}+v^2_{\xi;+}v^2_{\chi;0}-2v_{\xi;0}v_{\chi;+}v_{\xi;+}v_{\chi;0}\)\notag\\
&+&\f{\sigma_1}{2}v_\phi^2\(v_{\chi;+}^2+v_{\chi;0}^2\)+\f{\sigma_2}{4}v_\phi^2v_{\chi;+}^2+\f{\sigma_3}{2} \(v_{\xi;0}^2+2v_{\xi;+}^2\) v_\phi^2  \notag  \\
&+&\f{\sigma_4}{2\sqrt{2}}v_\phi^2\(v_{\chi;+} v_{\xi;+}+v_{\chi;0}v_{\xi;0}\).
\label{VEV:efpotential}
\eeqa
It should be noted that the couplings, such as $\la,\rho_i,\sigma_i$, etc., must be replaced with their values in regions of large field values evolved by RGEs. These evolved values can be utilized to determine the vacuum structure of the GM model. We disregard the dimensional parameters, which can be eliminated through the minimization conditions for the VEVs and play a negligible role in deriving the positive definiteness constraints in regions of large field values for the effective scalar potential.

It is evident that the VEVs of the custodial-preserving EWSB vacuum are significantly smaller than those of potential new vacua which may emerge in regions of large field values. As a consequence, the magnitude of the potential energy density for the custodial-preserving EWSB vacuum is considerably lower than that of the deeper vacua which arise at large field values, with the exception of certain highly exceptional cases.
To ensure vacuum stability, it is imperative that no lower-lying vacua, other than the conventional custodial-preserving vacuum, develop in regions of large field values. In other words, the scalar potential $V(v_\phi,v_\chi,v_\xi)$ must remain positive definite across all field-space asymptotes, analogous to the setting of ordinary vacuum stability bounds in the Standard Model.

The scalar potential $V(v_\phi,v_{\chi;+},v_{\chi;0},v_{\xi;+},v_{\xi;0})$ can be reduced to a matrix of quadratic form when the parameters $\rho_5,\sigma_4$ are sufficiently small, which takes the following form
\beqa
V=\left( v_\phi^2,v_{\chi;+}^2,v_{\chi;0}^2,v_{\xi;+}^2,v_{\xi;0}^2 \right)\left(\bea{ccccc}\f{\la}{4}&\f{2\sigma_1+\sigma_2}{8}&\f{\sigma_1}{4}&\f{\sigma_3}{2}&\f{\sigma_3}{4}\\\f{2\sigma_1+\sigma_2}{8}&\f{2\rho_1+\rho_2}{2}&\f{3\rho_2}{2}&\rho_4&\f{\rho_4}{2}\\\f{\sigma_1}{4}&\f{3\rho_2}{2}&\rho_1+\rho_2&\rho_4&\f{\rho_4}{2}\\\f{\sigma_3}{4}&\rho_4&{\rho_4}&2\rho_3&\rho_3\\\f{\sigma_3}{2}&\f{\rho_4}{2}&\f{\rho_4}{2}&\rho_3&\f{\rho_3}{2}\eea\right)\left(\bea{c} v_\phi^2\\v_{\chi;+}^2\\v_{\chi;0}^2\\v_{\xi;+}^2\\v_{\xi;0}^2 \eea\right).
\eeqa
A quadratic form is said to be positive definite if all of its eigenvalues are positive. According to Sylvester's criterion, this condition is equivalent to the requirement that all leading principal minors of the associated matrix are positive. To elaborate, this necessitates that
\beqa
&\lambda>0~, 
~~~\det\left(\bea{cc}\f{\la}{4}&\f{2\sigma_1+\sigma_2}{8}\\\f{2\sigma_1+\sigma_2}{8}&\f{2\rho_1+\rho_2}{2}\eea\right)>0~,~~~\det\left(\bea{ccc}\f{\la}{4}&\f{2\sigma_1+\sigma_2}{8}&\f{\sigma_1}{4}\\\f{2\sigma_1+\sigma_2}{8}&\f{2\rho_1+\rho_2}{2}&\f{3\rho_2}{2}\\\f{\sigma_1}{4}&\f{3\rho_2}{2}&\rho_1+\rho_2\eea\right)>0~,\notag\\
&\det\left(\bea{cccc}\f{\la}{4}&\f{2\sigma_1+\sigma_2}{8}&\f{\sigma_1}{4}&\f{\sigma_3}{2}\\\f{2\sigma_1+\sigma_2}{8}&\f{2\rho_1+\rho_2}{2}&\f{3\rho_2}{2}&\rho_4\\\f{\sigma_1}{4}&\f{3\rho_2}{2}&\rho_1+\rho_2&\rho_4\\\f{\sigma_3}{4}&\rho_4&{\rho_4}&2\rho_3\eea\right)>0~,~~~\det\left(\bea{ccccc}\f{\la}{4}&\f{2\sigma_1+\sigma_2}{8}&\f{\sigma_1}{4}&\f{\sigma_3}{2}&\f{\sigma_3}{4}\\\f{2\sigma_1+\sigma_2}{8}&\f{2\rho_1+\rho_2}{2}&\f{3\rho_2}{2}&\rho_4&\f{\rho_4}{2}\\\f{\sigma_1}{4}&\f{3\rho_2}{2}&\rho_1+\rho_2&\rho_4&\f{\rho_4}{2}\\\f{\sigma_3}{4}&\rho_4&{\rho_4}&2\rho_3&\rho_3\\\f{\sigma_3}{2}&\f{\rho_4}{2}&\f{\rho_4}{2}&\rho_3&\f{\rho_3}{2}\eea\right)>0~.\notag\\
\eeqa

%%%%%%%%%%%%%%%%%%%%%%
It should be noted that, while we have utilized the one-loop RG-improved tree-level scalar potential to numerically analyze the vacuum structure and the constraints of positive definiteness in the GM model, we have not considered the possibility that high-energy new physics could potentially destabilize the vacuum, leading to rapid vacuum decay with a tunneling time shorter than the age of the universe \cite{rvdthnp-1,rvdthnp-2,rvdthnp-3,rvdthnp-4,rvdthnp-5}. Moreover, the inclusion of gravitational effects at high energies could also significantly impact the decay rate of the false vacuum \cite{rvdthnp-5,rvdthnp-g-1,rvdthnp-g-2}. These aspects will be thoroughly investigated in our future works.

\end{itemize}

%%%%%%%%%%%%%%%%%%%%%%%%%%%%%%%%%%%%%%%%%%%%%%%
\section{\label{sec-5} Numerical results}

We would like to identify the parameters that can yield a light $CP$-even Higgs at 95 GeV to account for the excess signals reported by the ATLAS, CMS, and LEP collaborations. Realistic particle spectra, branching ratios, total widths of the scalars, and current experimental constraints, such as the mass of the SM-like Higgs, the couplings of the Higgs to vector bosons and fermions (cited in \cite{ATLAS:kappaWZ,CMS:kappaWZ,PDG:2022}), and B-meson rare decays sensitive to new physics (cited in \cite{B-physics}), can be calculated using the GMCALC package \cite{GMCal} (specifically designed for the Georgi-Machacek model) or the SPheno package \cite{SPheno1,SPheno2} (which, when combined with SARAH \cite{SARAH1,SARAH2,SARAH3,SARAH4,SARAH5}, can compute the spectrum of general new physics models). Given the rich phenomenology associated with additional Higgs bosons in the Georgi-Machacek model, we utilize the HiggsBounds-5.10.0 package \cite{HiggsTools,HiggsBounds1,HiggsBounds2,HiggsBounds3,HiggsBounds4} to ensure that our parameter choices comply with the bounds on additional Higgs bosons at colliders. Furthermore, we employ the HiggsSignals-2.6.2 package \cite{HiggsSignals1,HiggsSignals2,HiggsSignals3,HiggsTools} to verify that the next-to-lightest $CP$-even Higgs boson mimics the behavior of the SM-like Higgs.

Generally speaking, the input parameters include quadratic couplings $\lambda_{1,2,3,4,5}$, cubic coupling parameters with mass dimension $\mu_{1,2}$ and the mixing angle $\sin{\theta}$. Specifically, the input parameters $\mu_{1,2}$ can be alternatively expressed in terms of the singlet masses $m_h$ and $m_H$ to more easily facilitate the matching with their masses, as detailed in Eq. \ref{singlets-mass}. Besides In order to better control the degree of mixing between the two singlets, we can replace $\lambda_1$ with the mixing angle of two singlets $\alpha$ as elucidated by Eqs. (\ref{eq:singlets mass matrix} and \ref{eq:singlet mixing angle}).The range of these free input parameters is chosen as
\beqa
-\sqrt{4\pi}<{\lambda_{2,3,4,5}}<\sqrt{4\pi},~0<\sin{\theta}<0.45, \notag \\ 
~50~\mathrm{GeV} < m_{h,H} < 200~\mathrm{GeV},~0<\alpha<\pi.
\eeqa

Since our primary task is to investigate the specific impacts of the positive definiteness constraints on interpreting the 95 GeV excesses, we employ the numerical method provided in our recent work \cite{GM-pd} to check whether the parameter space that can explain these excess signals well fulfills the positive definiteness constraints or not.
The gauge coupling values at the EW scale used in this work are taken from Ref.~\cite{Antusch:2013jca}. The boundary conditions for the scalar quartic couplings at the EW scale, used in the process of RGE, are chosen according to Eq.~(\ref{rel}) to ensure that the vacuum preserves custodial symmetry. To determine the scale dependence of the quartic couplings in the one-loop renormalization group-improved tree-level scalar potential, it is necessary to evolve the input parameters to an arbitrary $\mu$ scale, particularly in regions of large field values, with the appropriate beta functions. In realistic numerical analyses, to establish positive definiteness constraints, we evolve the relevant quartic couplings and mass dimension parameters from the EW scale ($M_Z=91.2$ GeV) up to the Planck scale ($M_{Planck}\approx 1.12\times{10}^{19}$ GeV) or to the lowest Landau pole scale for the quartic couplings where any of them diverge.

%%%%%%%%%%%%%%%%%%%%%%%%%%%%%%%%%%%%%%%%%%%%%%%%%%%%%
\begin{figure}[htbp]
\begin{center}
\includegraphics[width=.49\textwidth]{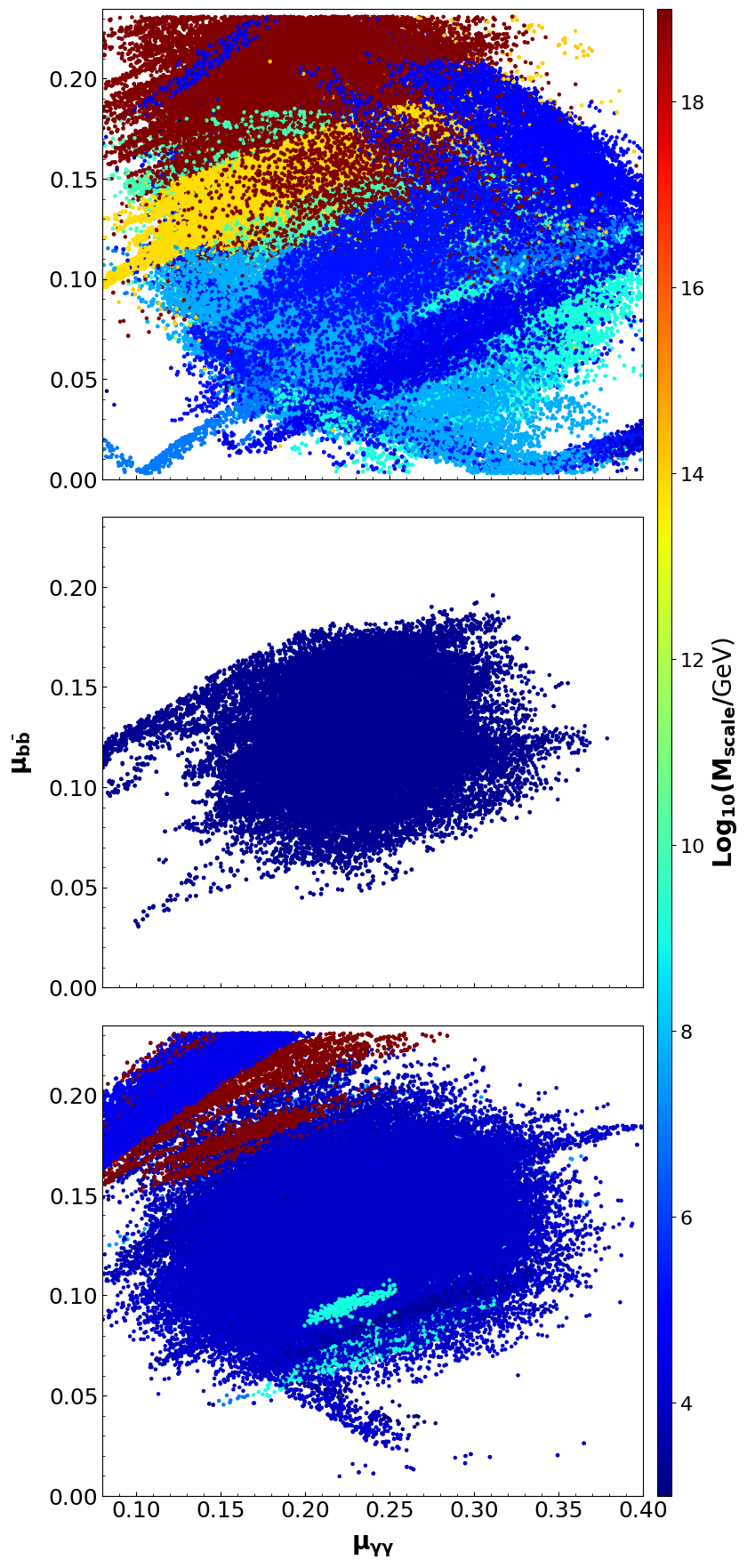}
\includegraphics[width=.49\textwidth]{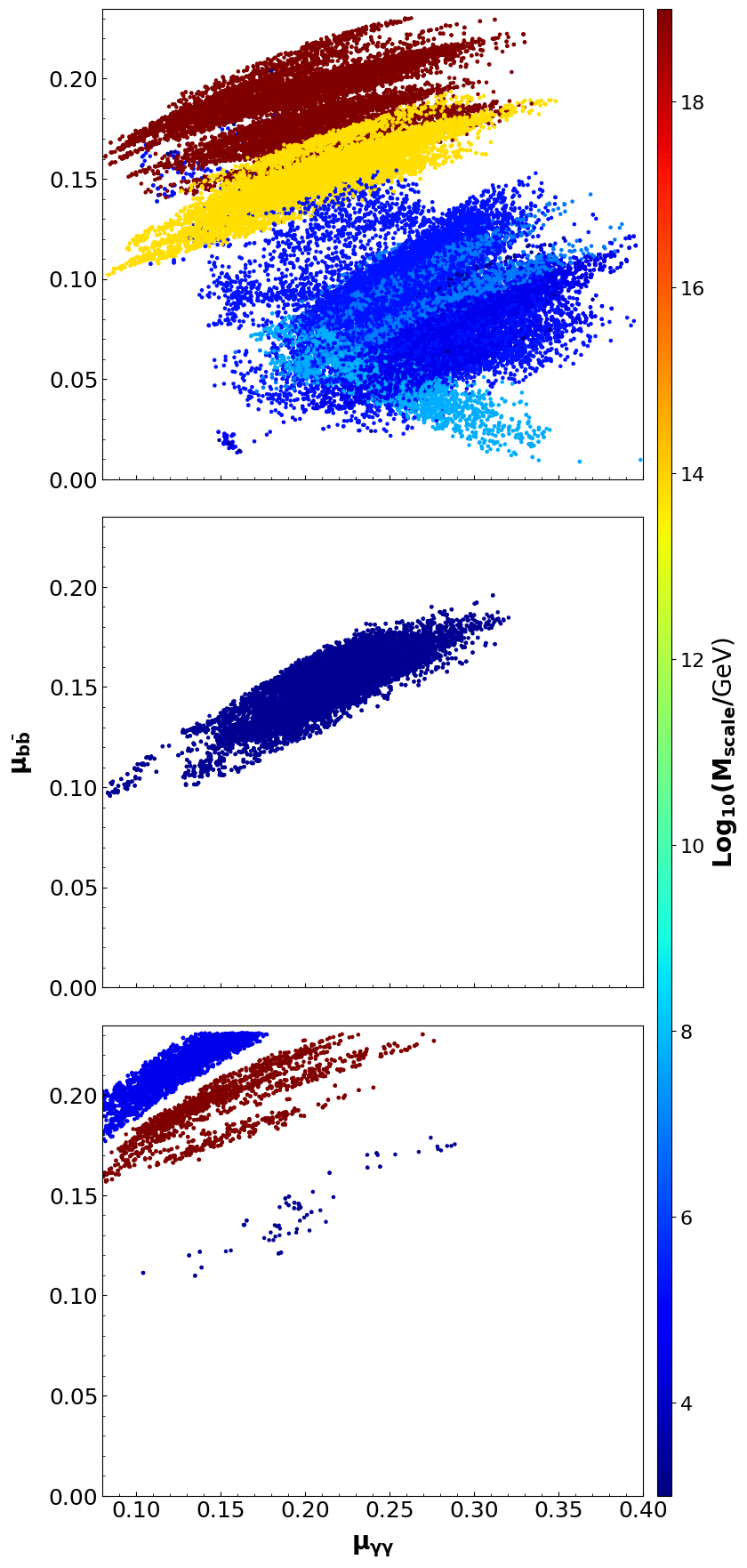}\vspace{-.5cm}
\end{center}
\caption{The signal strength of di-photon and $b \bar{b}$ excesses at 95 GeV. The different colors indicate the maximum values that each input parameter point can reach via RGE. All points in the top panels fulfill the positive definiteness constraints at large field value regions but do not meet the tree-level bounded-from-below bounds at the EW scale. The points in the middle two panels are completely opposite to those in the top. However, the panels in lower panels show points that satisfy both the tree-level bounded-from-below bounds at the EW scale and the positive definiteness constraints at large field value regions. All points in the left panels can successfully provide a 125 GeV SM-like Higgs boson and satisfy constraints of the theoretical perturbative unitarity bounds at the EW scale, B-meson rare decay and the oblique parameters `TSU'. Additionally, the right three figures are the result of considering the constraints related to the Higgs boson, as compared to the left panels.} 
\label{fig1} 
\end{figure}
%%%%%%%%%%%%%%%%%%%%%%%%%%%%%%%%%%%%%%%%%%%%%%%%%%%%%

We conduct a numerical scan and present our results in Fig. \ref{fig1}. Each set of input parameters must provide a next-to-lightest $CP$-even Higgs boson with a mass of 125 GeV, which serves as the SM-like Higgs boson, and a lightest $CP$-even Higgs boson to successfully explain the excess signals. We require that the points shown in Fig. \ref{fig1} must survive the theoretical perturbative unitarity bounds at the EW scale. Since the number of points obtained by the Markov Chain Monte Carlo (MCMC) method during the numerical scan is excessively large, we only display the points that can offer a quite good interpretation of the 95 GeV excess signals a $95\%$ confidence interval (corresponding to 2 $\sigma$).

The Fig. \ref{fig1} shows the sample distributions of signal strength for di-photon and $b \bar{b}$ excesses at 95 GeV provided by our work. And the different colors indicate the values of maximum scale that each input parameter point can reach via RGE.
The two panels at the top in Fig. \ref{fig1} (actually for all the upper panels on different plots in this work) show points that fulfill the positive definiteness constraints at large field value regions, but do not meet the tree-level bounded-from-below bounds at the EW scale. The points in the middle two panels are completely opposite to those on the top. The panels on the bottom, however, show points that satisfy both the tree-level bounded-from-below bounds at the EW scale and the positive definiteness constraints at large field value regions.
We observe that many parameter points, previously excluded by tree-level bounded-from-below constraints, can actually satisfy the positive definiteness constraints for the scalar potential when considering the one-loop RG-improved scalar potential at large field values regions. Conversely, a portion of the parameter space that meets the tree-level bounded-from-below constraints (using EW scale coupling inputs) should, in fact, be ruled out by the positive definiteness constraints of the effective potential at large field values. 
Indeed, this situation between the positive definiteness constraints at large field value regions and the tree-level bounded-from-below constraints is completely consistent with the conclusion of \cite{GM-pd}.

The sample points shown in the left three panels of Fig. \ref{fig1} can successfully provide a 125 GeV SM-like Higgs boson and satisfy the bounds of theoretical perturbative unitarity, B-meson rare decay and the oblique parameters ``TSU" at the EW scale, regardless of simultaneously explaining the 95 GeV excesses in 2 $\sigma$. It is evident that these two 95 GeV excesses can be well explained in the GM model, especially when considering the positive definiteness constraints in large field value regions and relaxing the tree-level bounded-from-below constraints. Due to the severe suppression of the interaction between the quark-phobic $SU(2)_V$ quintuple scalar and quarks, and the weak interaction between the gauge-phobic $SU(2)_V$ triplet scalar and fermions in the so-called ``alignment limit" of the GM model, the constraints of B-meson rare decay are not very strict. Moreover, considering the invariance of custodial symmetry at the tree level in the GM model, the constraints of the oblique parameters ``TSU" are also quite weak, which is also the benchmark feature of the GM model.
Additionally, the right three panels are the result of considering the constraints of extra Higgs bosons searching from collider data checked by the package of HiggsBounds-5.10 and the bounds of the specific properties of the SM-like Higgs boson, which is the heavier singlet scalar $h$ in this work, verified by the package of HiggsSignal-2.6.2. By comparing set of left panels, it becomes evident that the Higgs-related constraint has the most significant impact on the parameter space, particularly for parameter points that solely satisfy the tree-level vacuum stability constraint at the EW scale. Although many parameter points that satisfy positive definiteness bounds in the large field value regions have been excluded, the remaining points can still adequately explain the 95 GeV excess signals.

%%%%%%%%%%%%%%%%%%%%%%%%%%%%%%%%%%%%%%%%%%%%%%%%%%%%%
\begin{figure}[htbp]
\begin{center}
\includegraphics[width=.49\textwidth]{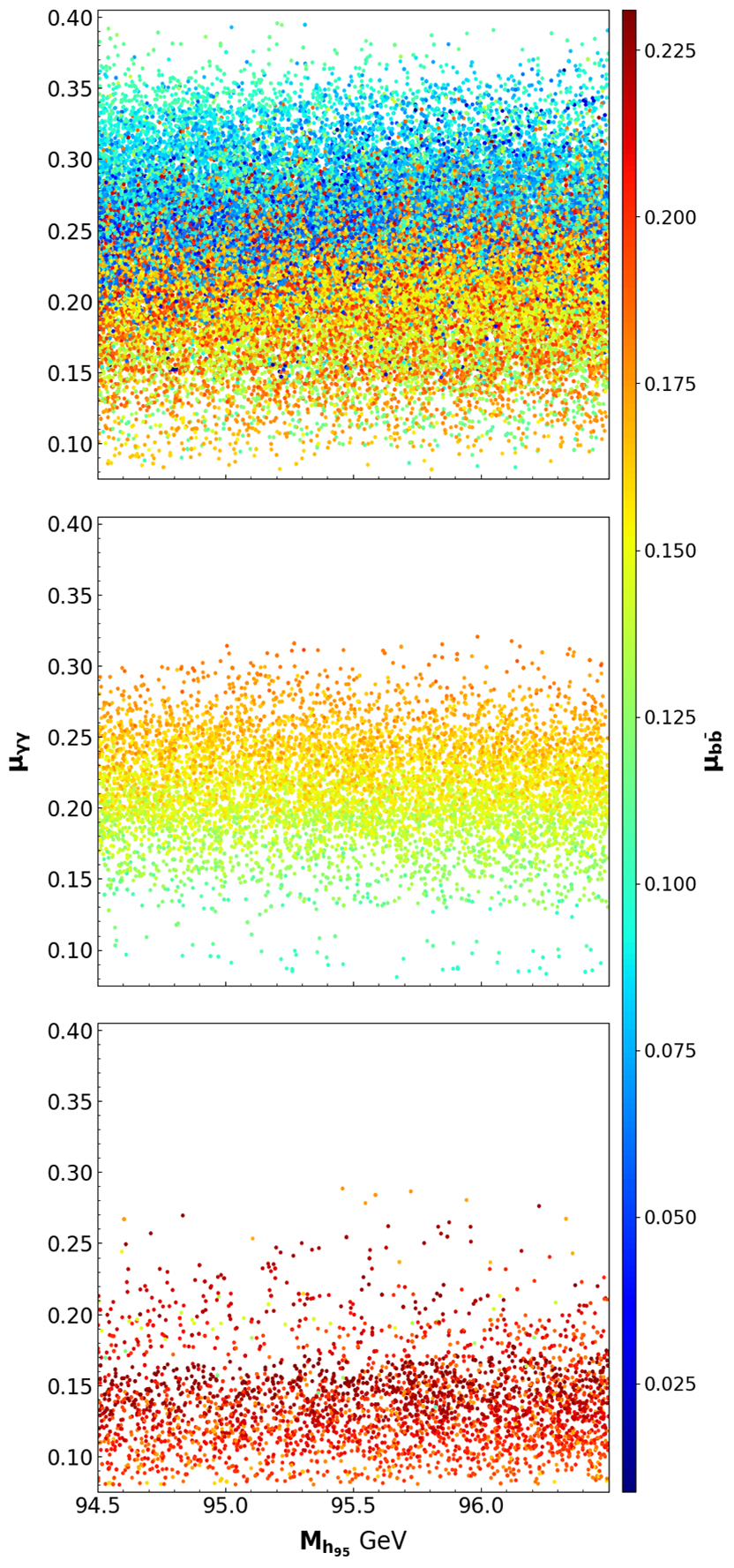}
\includegraphics[width=.49\textwidth]{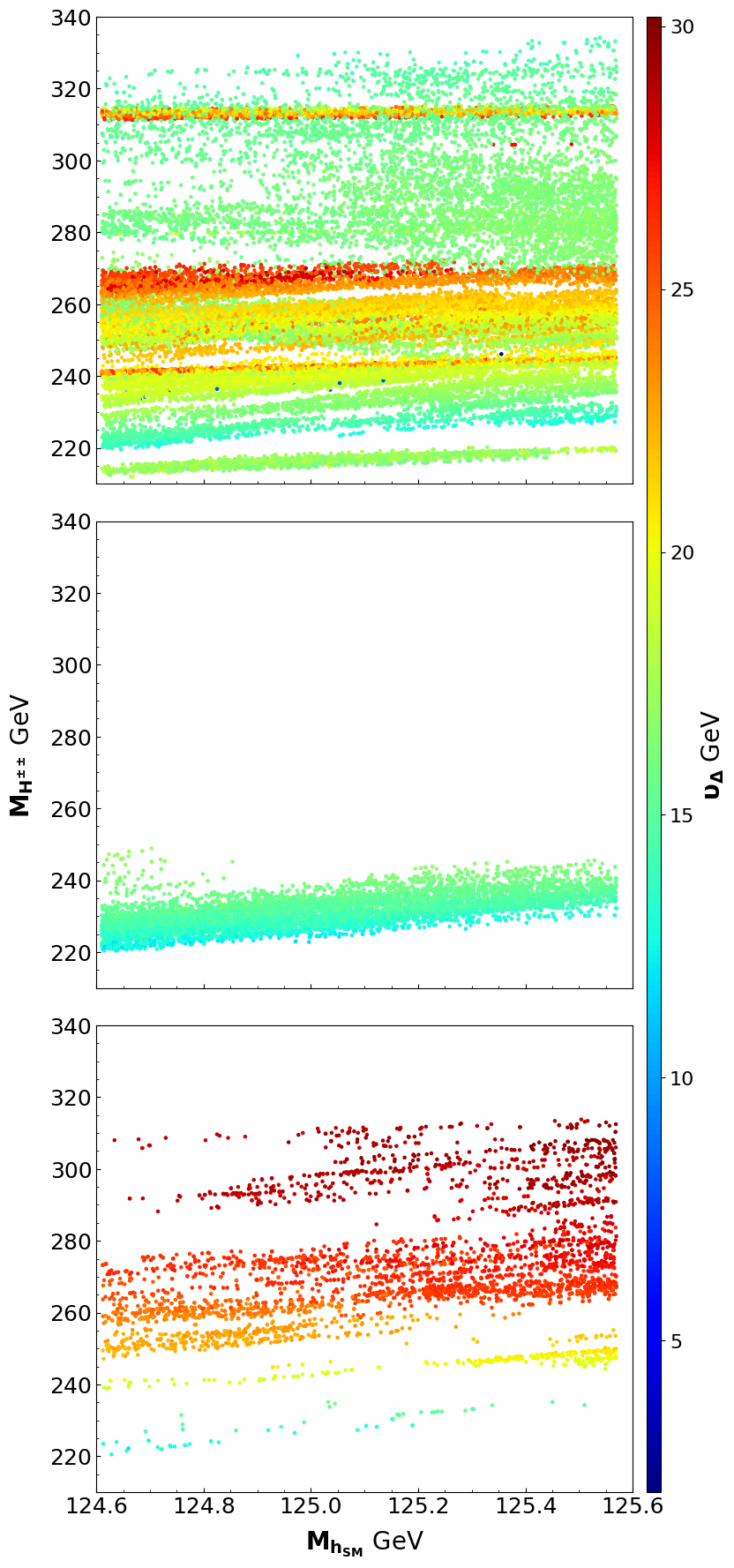}\vspace{-.5cm}
\end{center}
\caption{The left three panels of the figure illustrate the mass distribution of the light Higgs boson and its corresponding signal strength. The right three panels depict the mass ranges for both the SM-like Higgs boson and the doubly charged Higgs boson, with different colors indicating the VEVs of the triplet fields.}% The points presented here are detailed in the lower three panels of Figure \ref{fig1}. In other words, all points in this plot not only successfully account for a 125 GeV SM-like Higgs boson but also satisfy several critical constraints: the theoretical bounds of perturbative unitarity at the EW scale, the constraints related to B-meson rare decays, and the requirements imposed by the oblique parameters "TSU" as well as those associated with the Higgs boson itself.}
\label{fig2} 
\end{figure}
%%%%%%%%%%%%%%%%%%%%%%%%%%%%%%%%%%%%%%%%%%%%%%%%%%%%%

\begin{figure}[htbp]  
\begin{center}  
\includegraphics[width=.49\textwidth]{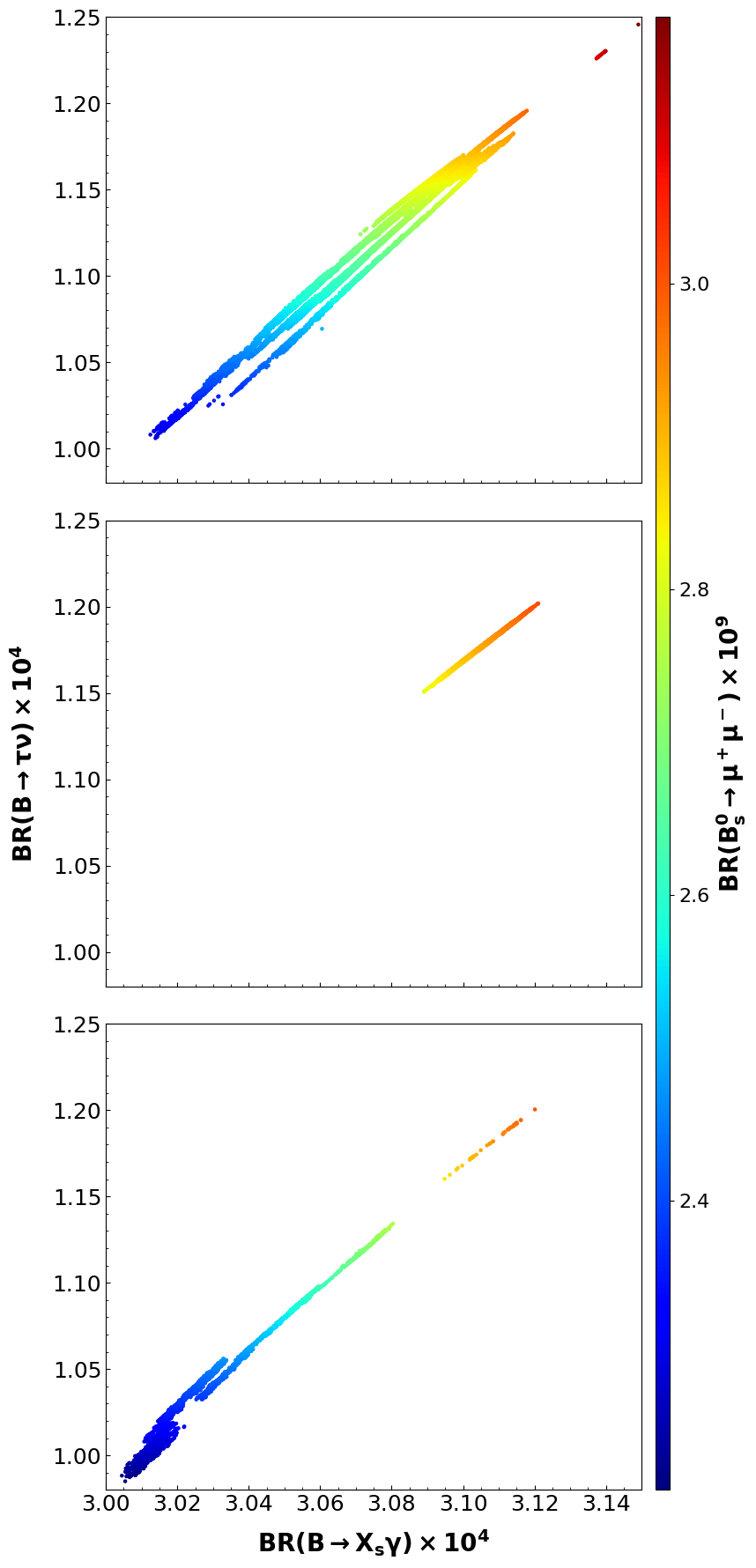}
\includegraphics[width=.49\textwidth]{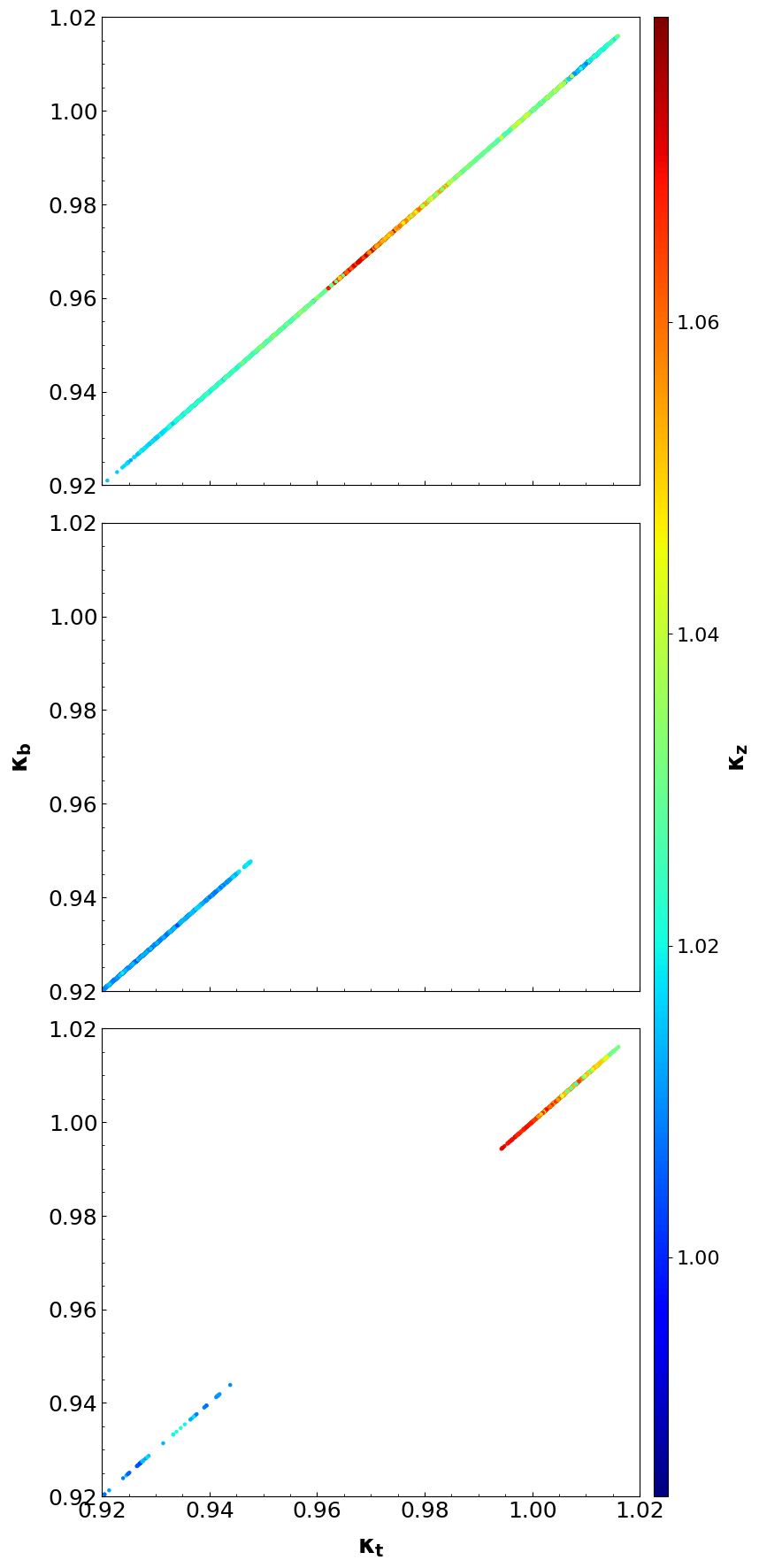}\vspace{-.5cm}
\end{center}
\caption{The left panels display the specific values of the branching ratios for the common B-meson rare decays in the $B\rightarrow\tau\nu$ channel, the $B\rightarrow {X_s \gamma}$ channel, and the $B_s^0\rightarrow\mu^+\mu^-$ channel. The values of the effective coupling for the interactions between the SM-like Higgs boson and third-generation quarks, as well as the Z vector boson are presented in the right panels.} 
\label{fig3}  
\end{figure}

After previous discussions, we can reach the important conclusion that the bounds of positive definiteness at large field value regions should be taken into account in the first place, rather than the constraints of bounded-from-below at the EW scale. To tell that more easily, we will conduct a specific analysis of the parameter space that satisfies the stringent theoretical and experimental constraints and can provide a very good explanation for the 95 GeV excess signals in the following part. Hence, all points in Fig. (\ref{fig2},\ref{fig3}) not only successfully account for a 125 GeV SM-like Higgs boson but also satisfy several critical constraints: the theoretical bounds of perturbative unitarity at EW scale, the constraints related to B-meson rare decays, and the requirements imposed by the oblique parameters ``TSU" as well as those associated with the Higgs boson itself. More precisely, the points presented in Fig. (\ref{fig2},\ref{fig3}) are detailed in the lower three panels of Fig. \ref{fig1}. Similar to Fig. \ref{fig1}, the points in the upper panels of Fig.(\ref{fig2},\ref{fig3}) fulfill the bounds of positive definiteness at large field value regions but can not satisfy the constraints of bounded-from-below at the EW scale, those in the middle panels are the opposite, and points in the lower panels can meet these two constraints. 

The left three panels of Fig. \ref{fig2} illustrate the mass distribution of the light Higgs boson with 95 GeV mass and its corresponding signal strength of $\mu_{\gamma\gamma}$ and $\mu_{b \bar{b}}$. 
It is easy to see that the parameter points which can provide a 95 GeV mass $CP$-even Higgs can offer good explanations for both di-photon and $b\bar{b}$ excesses. Of course, the parameter space which only satisfies the constraints of the positive definiteness at large field value regions is larger than those that only satisfies the bounds of bounded-from-below at EW scale as shown in the Fig. \ref{fig1}. 
What makes things particularly interesting is that the signal strength $\mu_{b\bar{b}}$ in the lower left panel, indicated by the points that can simultaneously satisfy both constraints, is quite large. Although these points can still explain the two excess signals within the 2$\sigma$ confidence region, the reason that the points with smaller signal strengths $\mu_{b\bar{b}}$ have been excluded is due to bounds on extra Higgs bosons searches at colliders and the specific properties of the SM-like Higgs.

The right three panels of Fig. \ref{fig2} depict the mass ranges for both the SM-like Higgs boson and the doubly charged Higgs boson, with different colors indicating the VEVs of the triplet fields. For points fulfilling the constraints of positive definiteness in large field value regions (shown in the upper right and the lower right panels), the mass of the doubly charged Higgs boson varies from 200 GeV to 340 GeV, whereas for those satisfying the bounded-from-below conditions at the EW scale (shown in the middle right panel), it ranges only from 220 GeV to 250 GeV; the narrow mass region is consistent with the tendency of $H_5$ provided by \cite{95-38}.
The mass of the doubly charged Higgs slowly increases with that of the SM-like Higgs and shows a significant positive correlation with the VEVs of the triplets which varies from 3 GeV to 30 GeV. 
%Interestingly, in the lower left panel, there exists an independent parameter space that can provide a wide mass range for the doubly charged Higgs, particularly when the VEVs of the triplets are around 16 GeV.

%%%%%%%%%%%%%%%%%%%%%%%%%%%%%%%%%%%%%%%%%%%%%%%%%%%%%
%%%%%%%%%%%%%%%%%%%%%%%%%%%%%%%%%%%%%%%%%%%%%%%%%%%%%

All the points in Fig. \ref{fig3} have taken into account the constraints from B-meson rare decays, oblique parameters ``TSU" and the effective couplings of the SM-like Higgs boson. So it is clear that all the parameter space shown in Fig. \ref{fig3} can survive well under the experimental constraints from the common B-meson rare decays (in the $B\rightarrow\tau\nu$ channel, the $B\rightarrow {X_s \gamma}$ channel, and the $B_s^0\rightarrow\mu^+\mu^-$ channel) and the effective coupling of the SM-like Higgs boson (the interactions between the SM-like Higgs boson and the top quark $\kappa_t$, bottom quark $\kappa_b$, and the Z vector boson $\kappa_{ZZ}$). 

Thanks to the suppressed interaction of the quark-phobic $SU(2)_V$ quintuplet scalar with quarks and the weakly coupled nature of the gauge-phobic $SU(2)_V$ triplet scalar with fermions in the ``alignment limit" of the GM model, constraints from B-meson rare decays are relatively weak even when we apply the newest data \cite{PDG:2024}, as demonstrated in the left three panels of Fig. \ref{fig3}. Consequently, most of the parameter space can survive the incoming collider experiments, such as the High-Luminosity LHC (HL-LHC)~\cite{HL-LHC1,HL-LHC2}.

On the contrary, constraints from the precision measurements of Higgs coupling are very stringent for the GM model, which extends the SM with triplet scalar fields. Although the points in the right three panels of Fig. \ref{fig3} currently meet all bounds, the parameter space will shrink significantly once future collider experiments are considered. In particular, applying the predicted bounds from the Circular Electron-Positron Collider (CEPC) ~\cite{CEPC1,CEPC2,CEPC3} (with 0.01 accuracy for top and bottom quarks and 0.001 accuracy for Z bosons), both the lower middle and right panels would be entirely excluded, while only a minimal portion of the parameter space would remain viable for the upper right panel.

%%%%%%%%%%%%%%%%%%%%%%%%%%%%%%%%%%%%%%%%%%%%%%%
\begin{table}[htbp]  
\centering  
%%%%%%%%%%%%%%%%%%%%%%%%%%%%%%%%%%%%%%%%%%%%%%%%%%%%%%%%%%%%%%%%
% first table  
\begin{tabular}{cccccc}  
%\toprule % 顶部粗线 
\multicolumn{6}{c}{EW scale}   \\ 
\hline 
$\lambda_1$ & 0.0335~~~0.0320       & $\lambda$ & 0.1340~~~0.1278       & $\sigma_4$ & 0.7862~~~0.0417  \\  
$\lambda_2$ & -0.9263~~~-0.2818     & $\rho_1$ & 0.3734~~~-3.1750       & $\mu^\prime_1$ & 44.3818~~~30.3390  \\  
$\lambda_3$ & 0.6798~~~-0.3413      & $\rho_2$ & -2.7191~~~1.3651       & $\mu^\prime_2$ & 31.3827~~~21.4529   \\  
$\lambda_4$ & 0.2608~~~0.2597       & $\rho_3$ & -0.4931~~~-1.2463      & $\mu^\prime_3$ & -3937.8724~~~-1210.3213  \\  
$\lambda_5$ & -0.5559~~~-0.0295     & $\rho_4$ & -1.8526~~~-0.5637      & $m_{\phi}$ & 58.8545~~~83.2217  \\  
$m_h$ & 126.8221~~~128.2429         & $\rho_5$ & 2.7191~~~-1.3651       & $m_\xi$ & 343.1990~~~113.1068   \\  
$m_H$ & 121.9700~~~120.6273         & $\sigma_1$ & 0.4871~~~1.0092      & $m_\chi$ & 355.0436~~~111.2011   \\  
$\sin(\theta_H)$ & 0.5069~~~0.2067  & $\sigma_2$ & 1.1112~~~0.0589      & $\upsilon_{\Delta}$ & 15.1016~~~17.9776   \\  
$\alpha$ & 0.1736~~~1.9891          & $\sigma_3$ & 0.5215~~~0.5193     & $\upsilon_{\phi}$ & 243.3934~~~242.1265  \\  
\hline   
\end{tabular}  

\vspace{0.2cm}
%\bigskip  

% third table  
\begin{tabular}{cccccc}  
%\toprule % 顶部粗线 
\multicolumn{6}{c}{Observeables}   \\ 
\hline 
$m_{h_1}~(m_H)$ & 95.5927~~~96.1130    & $\kappa^2_{Hff}$   & 0.3846~~~0.1947        &  $\kappa_{b}$ & 0.9398~~~1.0034 \\  
$m_{h_2}~(m_h)$ & 125.5482~~~125.3452  & $BR(H\to\ga\ga)$   & 0.0009~~~0.0017        & $\kappa_{\tau}$ & 0.9398~~~1.0034  \\  
$m_{h_3}$ & 368.3684~~~253.7774        & $BR(H\to b\bar{b})$  & 0.8004~~~0.7980      & $\kappa_{ww}$ & 1.0185~~~1.0247 \\  
$m_{a_2}$ & 310.0807~~~185.8671        & $BR(H\to\tau\bar{\tau})$  & 0.0887~~~0.0885 & $\kappa_{zz}$ & 1.0193~~~1.0249 \\  
$m_{H^\pm_2}$ & 237.4317~~~186.1513    & $\mu_{\ga\ga}$   & 0.2490~~~0.2381          & S   &  0.0330~~~-0.0795 \\  
$m_{H^\pm_3}$ & 323.3204~~~251.3946    & $\mu_{b\bar{b}}$   & 0.1115~~~0.1436        & T   & -0.1027~~~-0.0164 \\  
$m_{H^{\pm\pm}}$ & 324.9837~~~252.1293 & $\mu_{\tau\bar{\tau}}$   & 0.1190~~~0.1533  & U   & -0.0200~~~0.0005 \\  
$\kappa^2_{HVV}$ & 0.1116~~~0.1441    & $\kappa_{t}$ & 0.9398~~~1.0034               & P-value   & 0.8125~~~0.8100 \\  
\hline   
\end{tabular}  

\vspace{0.2cm}
%\bigskip  
  
% fourth table    
\begin{tabular}{cccc}  
\multicolumn{4}{c}{Observeables}  \\ 
\hline  
$M_{scale}$ & $3.1623\times{10}^6~~~6.3096\times{10}^{16}$                              & HiggsSignals & $\checkmark~~~\checkmark$ \\  
$Br(B_s\rightarrow X_{s}\gamma)$ & $3.1086\times{10}^{-4}~~~3.0901\times{10}^{-4}$      & Unitarity & $\checkmark~~~\checkmark$ \\  
$Br(B^0_s\rightarrow \mu^{+}\mu^{-})$ & $2.8925\times{10}^{-9}~~~2.7844\times{10}^{-9}$   & VS at EW & $\times~~~\times$ \\  
$Br(B\rightarrow \tau\nu)$ & $1.1725\times{10}^{-4}~~~1.1431\times{10}^{-4}$    & VS at High scale & $\checkmark~~~\checkmark$ \\  
HiggsBounds & $\checkmark~~~~~~~~~~~~~~~~~~\checkmark$                                                 & Positive Definiteness & $\checkmark~~~\checkmark$ \\ 
\hline  
\end{tabular}  
%%%%%%%%%%%%%%%%%%
\caption{Two benchmark points. The first sub-table shows the input parameters in (\ref{eq:GMpot}) and the values of corresponding parameter in the scalar potential (\ref{eq:GMpot_full}) at the EW scale. We also present some low energy experimental observables in the second and the third sub-tables. All dimensional parameters listed in the table are given in units of GeV.}
\label{tab:Benchmark}
\end{table}  
%%%%%%%%%%%%%%%%%%%%%%%%%%%%%%%%%%%%%%%%%%%%%%%%%%%%%
We present two benchmark points, which can satisfy both current experimental and theoretical constraints except the bounded-from-below constraints adopted in previous studies while fulfill the positive definiteness constranints in the large field value regions, in Table \ref{tab:Benchmark} that can adress these two excesses of di-photon and $b\bar{b}$.

%%%%%%%%%%%%%%%%%%%%%%%%%%%%%%%%%%%%%%%%%%%%%%%%%%%%%
%%%%%%%%%%%%%%%%%%%%%%%%%%%%%%%%%%%%%%%%%%%%%%%%%%%%%

\section{\label{sec-6} Conclusions}

In this work, we investigated whether the GM model can accommodate the 95 GeV excesses observed in the di-photon and $b\bar{b}$ channels while satisfying theoretical and experimental constraints. By incorporating the newly proposed positive definiteness conditions, we performed a detailed numerical analysis and found that the GM model remains a viable explanation for these excesses. More importantly, our results reveal that these stronger vacuum stability constraints, specifically the bounds of positive definiteness at large value field regions, not only allow for parameter space where a 95 GeV custodial singlet Higgs can explain the observed excesses, but also expand the allowed parameter regions compared to the traditional tree-level bounded-from-below conditions at the EW scale. This enhancement significantly improves the model’s viability.

Furthermore, we emphasize the crucial role of future collider experiments, such as HL-LHC and CEPC, in rigorously testing this interpretation. These experiments will refine Higgs coupling measurements, probe additional Higgs bosons, and directly test key predictions of the GM model. The increased precision and discovery potential of these facilities will be essential in determining whether the GM model can truly explain the 95 GeV excesses or if alternative scenarios should be considered.

%%%%%%%%%%%%%%%%%%%%%%%%%%%%%%%%%%%%%%%%%%%%%%%%%%%%%
%%%%%%%%%%%%%%%%%%%%%%%%%%%%%%%%%%%%%%%%%%%%%%%%%%%%%

\begin{acknowledgments}
We would like to thank useful discussions with Fei Wang and Yang Zhang. This work was supported by the National Natural Science Foundation of China (Grant Nos. 12447167 and 12275067), by the Joint Fund of Henan Province Science and Technology R$\&$D Program (Grant Nos.225200810030, 225200810092 and 225200810014), by the Startup Research Fund of Henan Academy of Sciences (Grant No. 231820011), by the Basic Research Fund of Henan Academy of Sciences (Grant No. 240620006), by the Graduate Innovation Fund of Henan Academy of Sciences (Grant No. 243320031), by the Science and Technology Innovation Leading Talent Support Program of Henan Province, by the National Key R$\&$D Program of China (Grant No. 2023YFA1606000) and by the Science and Technology Innovation Leading Talent Support Program of
Henan Province  (Grant No. 254000510039).
\end{acknowledgments}

%%%%%%%%%%%%%%%  reference

\end{document}